# Imaging Transformer for MRI Denoising: a Scalable Model Architecture that enables $SNR \ll 1$ Imaging

**Hui Xue[1], Sarah M. Hooper[2], Rhodri H. Davies[3,4], Thomas A. Treibel[4], Iain Pierce[4], John Stairs[1], Joseph Naegele[1], Charlotte Manisty[4], James C. Moon[4], Adrienne E. Campbell-Washburn[2], Peter Kellman[1]*, Michael S. Hansen[1]***

1. Microsoft Research, Health Futures, Redmond, WA, USA
2. National Heart, Lung and Blood Institute, National Institutes of Health, Bethesda, MD, USA
3. Institute of Cardiovascular Science, University College London, London, UK
4. Barts Heart Centre, Barts Health NHS Trust, London, UK

*These authors contributed equally as senior authors to this work.

## Corresponding author:

Hui Xue

Microsoft Research, Health Futures

Building 99, Room 4941
14820 NE 36th St
Redmond
WA 98052

Email: xueh@microsoft.com

***This manuscript is currently submitted to Radiology: Artificial Intelligence for consideration to be published as a research paper.***

# Imaging Transformer for MRI Denoising: a Scalable Model Architecture that enables $SNR \ll 1$ Imaging

**Key points:**

1. We introduced a novel and flexible imaging transformer architecture for MRI denoising. The key innovation lies in decomposing full attention across imaging data into intra-frame local, global, and inter-frame attention computations.

2. The imaging transformer model recovered image quality sufficient for confident clinical readings and accurate measurement of ejection fraction (EF) even for inputs with an SNR of 0.2, where images were uninterpretable without denoising.

3. Effects of scalability were experimentally verified with performance improved as model size increases. The larger model had more accurate EF measurement than smaller ones. Training was performed on a large dataset of 206,677 cine series and testing was on 7,267 samples. The imaging transformer outperformed all seven convolutional and transformer baselines.

**Summary statement:**

We present a novel imaging transformer that offers superior performance, scalability, and versatility for MRI denoising. It can recover high-quality images suitable for confident clinical reading, even at very low input SNR levels of 0.2.

**Abbreviations**

magnetic resonance imaging = MRI, signal-to-noise ratio = SNR, convolutional neural network = CNN, vision transformer = ViT, shifted window transformer = Swin, Imaging Transformer = IT, high-resolution network = HRNet , ejection fraction = EF, ground-truth = GT, balanced steady-state free precession = B-SSFP, left ventricular = LV, two-chamber = CH2, three-chamber = CH3, four-chamber = CH4, short-axis stack = SAX, standard deviation = SD, peak signal-noise-ratio = PSNR, structural similarity index measure = SSIM, contrast-to-noise ratio = CNR, confidence range = CR

## Abstract

### Purpose

To propose a flexible and scalable imaging transformer (IT) architecture with three attention modules for multi-dimensional imaging data and apply it to MRI denoising with very low input SNR.

### Methods

Three independent attention modules were developed: spatial local, spatial global, and frame attentions. They capture long-range signal correlation and bring back the locality of information in images. An attention-cell-block design processes 5D tensors ([B, C, F, H, W]) for 2D, 2D+T, and 3D image data. A High Resolution (HRNet) backbone was built to hold IT blocks. Training dataset consists of 206,677 cine series and test datasets had 7,267 series. Ten input SNR levels from 0.05 to 8.0 were tested. IT models were compared to seven convolutional and transformer baselines. To test scalability, four IT models 27m to 218m parameters were trained. Two senior cardiologists reviewed IT model outputs from which the EF was measured and compared against the ground-truth.

### Results

IT models significantly outperformed other models over the tested SNR levels. The performance gap was most prominent at low SNR levels. The IT-218m model had the highest SSIM and PSNR, restoring good image quality and anatomical details even at SNR 0.2. Two experts agreed at this SNR or above, the IT model output gave the same clinical interpretation as the ground-truth. The model produced images that had accurate EF measurements compared to ground-truth values.

### Conclusions

Imaging transformer model offers strong performance, scalability, and versatility for MR denoising. It recovers image quality suitable for confident clinical reading and accurate EF measurement, even at very low input SNR of 0.2.

## Introduction

MRI denoising aims to recover signals from low signal-to-noise ratio (SNR) acquisitions. Ideally, MRI should provide high SNR, resolution, and contrast to visualize anatomical structures for disease identification. However, imaging physics imposes trade-offs—SNR decreases with smaller pixel size and shorter acquisition times, such as in undersampled parallel imaging.

Deep neural networks have emerged as powerful denoisers (1–12), enabling faster acquisitions with improved diagnostic accuracy (4,9,13). Most published work exploited variants of convolutional neural networks (CNN) (5,6,8,10–12,14). Examples include Resnet (15) to denoise the diffusion weighted brain scans, feed-forward CNNs to reduce noise in neuro T1, T2 and fluid attenuated 3D MRI (11), and U-net to enhance knee MRI (10). These CNN solutions significantly improved over traditional filtering-based methods (2).

Recent advances in computer vision favor transformer-based architectures (4,16,17), which outperform CNNs in classification, detection, and segmentation tasks by capturing long-range dependencies and adapting to input variations (2,18–22). The reason can be attributed to a few fundamental differences between the architectures. First, the attention coefficients in transformers are computed on-the-fly for every input sample, potentially enabling them to adapt better, compared to convolutional kernels' fixed parameters after training. Second, transformers are better at capturing long-range dependencies between signals and enable larger effective receptive fields, experimentally consistent with improved performance (23).

Originally developed for language modeling (24), transformers face challenges in imaging due to the quadratic complexity of attention mechanisms, making full-image attention computationally expensive. Vision Transformer (ViT) (25) introduced image tokenization, splitting the whole volume into patches and computing attention over tokens rather than pixels, surpassing CNN baselines (26). However, its full-attention design is computationally expensive and lacks inductive bias for locality, which is crucial for imaging where neighboring pixels are highly correlated. The Shifted Window Transformer (Swin) (21) addressed this by computing attention within local windows overlapped with each other by shifting, reintroducing some inductive bias. Originally designed for classification and segmentation, Swin has been adapted for denoising (22), proving effective for 3D arterial spin labeling MRI, outperforming ViT and ResNet baselines (4).

Both ViT and Swin simply stacked their transformer modules into a feed-forward backbone. As the denoising task requires pixel-wise prediction on the original image resolution, this design does not take advantage of multi-resolution scheme to balance computation and model capacity (27). Other studies proposed to replace convolution modules in a U-net architecture by transformer modules (19), to outperform feed-forward backbone.

A good imaging architecture enables global pixel interactions to capture long-range correlations while benefiting from locality and reducing computational complexity. Versatility should be provided to handle diverse data formats, such as 2D, 2D+T or 3D (there are limited sets of 3D+T scenarios which are not targeted here). We propose a new set of transformer modules that are collectively termed *Imaging Transformer (IT)*, applied to MRI denoising. It was designed to decouple intra-frame local, global, and inter-frame correlations. These modules process images flexibly, taking a 5D tensor [B, C, F, H, W] as input and output. Here, B is batch size, C is channels, H and W are height and width, and F represents time or depth or slice in 3D volumes. For 2D imaging, F is unitary.

Our model effectively recovers signals from extremely low SNR inputs (Figure 1a). By decomposing full attention into local (L), global (G), and frame (F) modules, we reduce computational complexity to linear scaling with the size of data volume. Each module operates independently, allowing customization for capacity and flexibility by inserting additional IT blocks. This design provides more explicit control to where the model should pay more attention. For example, more frame modules can be inserted into the block to enhance inter-frame modelling. In this study, we trained IT-based denoising networks ranging from 27M to 218M parameters on a high-resolution network (HRNet (28), Figure 1d) backbone using 206,677 2D+T cardiac cine series and tested on 7,267. The IT models were compared with other baselines. Model outputs were evaluated by two expert cardiologists for diagnostic confidence. The EF from model images were compared to the ground-truth measurements.

## Materials and Methods

### *Data collection*

Data for this retrospective study was from the NIH Cardiac MRI Raw Data Repository, hosted by the Intramural Research Program of the National Heart, Lung and Blood Institute. All data were curated with the required ethical and/or secondary audit use approvals and guidelines that permitted retrospective analysis of anonymized data without requiring written informed consent for secondary usage for the purpose of technical development, protocol optimization, and quality control. All data were fully anonymized and used in training without exclusion.

The training data consists of breath-held retro-gated cardiac cine imaging from 3T clinical scanners (MAGNETOM Prisma, Siemens AG Healthcare). A dataset of 206,677 cine series (6,160,700 frames) from 16,220 patients was compiled. Imaging used a balanced steady-state free precession (B-SSFP) sequence with typical parameters as the following: field-of-view 360×270 mm², matrix 256×144, bandwidth 977 Hz/pixel, flip angle 50°, parallel acceleration R=2, echo spacing 2.97ms, echo time 1.28ms, and 30 phases over 7-10 heartbeats. Scans included two-chamber (CH2), three-chamber (CH3), four-chamber (CH4), and short-axis stack (SAX) views. Raw k-space signals were stored for reconstruction. The test set contained 7,267 series from 525 patients, with no overlap between training and test data.

***Imaging Transformer (IT) modules***

IT modules process a 5D tensor and output another 5D tensor by computing attentions along spatial and frame dimensions. Given the input tensor $\boldsymbol{x}$, the Query (Q), Key (K), and Value (V) tensors are first computed. To introduce locality bias and support variable matrix sizes, three convolutions are used to compute Q, K, and V. These convolution kernels are learnable parameters, differing from standard attention that uses linear layers. This design allows freely adjusting the number of channels through the network. As shown later, the channel is proportional to dimensions of **q/k/v** tensors which controls the capacity of imaging attention. That is, models can be scaled up by increasing number of channels as well as adding more attention modules.

*Spatial local attention (L)*

With the Q/K/V tensors computed, they are divided into attention windows containing multiple patches. Given the window size $[w, w]$ and patch size $[p, p]$, the number of patches per window is $P = \frac{w}{p} \cdot \frac{w}{p}$. The name "spatial local" implies the attention computation is limited to all patches within one attention window on every $[H, W]$ 2D frame. Every patch has $p^2$ pixels. Flattening all pixels in a patch across the $\boldsymbol{C}$ channel gives the pixel vector $\boldsymbol{v} \epsilon \mathbb{R}^{Cp^2 \times 1}$. Stacking all $P$ vectors as a data matrix $\boldsymbol{D}$ gives:

$$\boldsymbol{D}_{local} = \begin{bmatrix} \boldsymbol{v}_0^T \\ \boldsymbol{v}_1^T \\ \vdots \\ \boldsymbol{v}_{P-1}^T \end{bmatrix} \in \mathbb{R}^{P \times Cp^2}$$

These above-mentioned steps were repeated for $\boldsymbol{Q}$, $\boldsymbol{K}$ and $\boldsymbol{V}$, resulting in $\boldsymbol{D_Q}$, $\boldsymbol{D_K}$ and $\boldsymbol{D_V}$. A larger $\boldsymbol{C}$ increases attention capacity which allows more information to be stored in $\boldsymbol{K}$ and $\boldsymbol{V}$. The attention matrix is computed between $\boldsymbol{D_Q}$ and $\boldsymbol{D_K}$: $\boldsymbol{A}(\boldsymbol{D_Q}, \boldsymbol{D_K}) = softmax(\boldsymbol{D_Q}\boldsymbol{D_K^T}/\sqrt{Cp^2} + \boldsymbol{B})$. $\boldsymbol{A}$ is the $\mathbb{R}^{P \times P}$attention coefficient matrix. $\boldsymbol{B}$ is the relative positional bias matrix (29). The output for this window is $\boldsymbol{A} \cdot \boldsymbol{D_V}$. This attention computation is performed for every window with the batched matrix computation. The final output is reshaped back to be a 5D tensor. A multi-head attention version is achieved by splitting $\boldsymbol{v}$ for every head and computing attention matrixes for all heads.

*Spatial global attention (G)*

Global attention captures dependencies between remote regions, while the local attention focuses on neighboring patches. As shown in Figure 1c, global attention assembles the data matrix over corresponding patches from all attention windows. Given the window size $[w, w]$, the number of attention windows is $N = \frac{H}{w} \cdot \frac{W}{w}$. Stacking patches gives the global attention data matrix $\boldsymbol{D}$:

$$\boldsymbol{D}_{global} = \begin{bmatrix} \boldsymbol{v}_0^T \\ \boldsymbol{v}_1^T \\ \vdots \\ \boldsymbol{v}_{N-1}^T \end{bmatrix} \in \mathbb{R}^{N \times Cp^2}$$

The attention computation is over the same color patches from $N$ windows (Figure 1c), resulting the $\mathbb{R}^{N \times N}$attention coefficient matrix which is applied to the *value* tensor.

We also experimented with random shuffling the patches in a window, then computed attention. But this extra step had no impact on model performance.

*Frame attention (F)*

Correlations between frames in imaging are strong. Processing every 2D frame independently is not optimal. The number of frames in a scan is in general much less than the number of pixels. It is feasible to perform full attention computation along the frame dimension. Given the 5D tensor with $M$ frame, the data matrix is assembled for all frames:

$$\boldsymbol{D}_T = \begin{bmatrix} \boldsymbol{v}_0^T \\ \boldsymbol{v}_1^T \\ \vdots \\ \boldsymbol{v}_{M-1}^T \end{bmatrix} \in \mathbb{R}^{M \times CHW}$$

$\boldsymbol{v_i} \epsilon \mathbb{R}^{CHW \times 1}$ contains all pixels in the *i-th* 2D frame flattened. The resulting attention coefficient $\boldsymbol{A}$ is a $M \times M$ matrix. All output frames are computed by multiplying $\boldsymbol{A}$ to the *value* tensor.

***Backbone and training***

As shown in Figure 1d, following the classical design (24), we integrate attention modules with other layers into a Cell which consists of an attention module, layer normalizations (30), skip connections (15), and a mixer. The attention module can be G, L, or F. The mixer includes layer normalization, PReLU activation (31), and two convolution layers: the first scales the channels by a factor of 4, and the second restores the original channel count. Multiple cells form a processing Block, and we use acronyms G, L, or F to indicate the attention module type in the block. For example, a FLG block consists of frame, local, and global attention cells. The model can be scaled by adding cells; for instance, a FLGFLG block has six cells, doubled the size. A dropout (32) of 0.1 is applied to cell outputs for regularization. All convolution kernels are 3×3 with padding of 1.

Attention modules, cells, and blocks form the three levels of building blocks for the full imaging transformer model. We use the HRNet backbone, having more processing blocks on the highest resolution representations by avoiding early-stage downsampling (Figure 1e). For imaging where fine-grained details are crucial, this architecture favors signal recovery with high fidelity. We replace the convolution layers of the original HRNet CNN with imaging transformer blocks.

The SNRAware training method (33) synthesizes training samples from high-SNR data, adding MR imaging noise augmented with g-factor maps of varying accelerations. It normalizes the noise level to unity through SNRUnit (34) reconstruction instead of normalizing signal levels (35). Inputs include complex image series and g-factor maps, and outputs are denoised complex images. The input channel $C_{in}$ is 3 (real and complex values with the g-factor map), and the output channel $C_{out}$ is 2. The encoding channel C is 64. The training set uses 95% for training and 5% for validation. Training setup is as follows: the Sophia optimizer (36), a one-cycle learning rate scheduler (37), peak learning rate 1e-5, betas 0.9 and 0.999, epsilon 1e-8, 160 epochs, Pytorch 2.6 (38), a cluster of 128 AMD MI300X GPUs, each with 192GB RAM. Further details are in Appendix E1 (supplement).

### *Evaluation*

A large test dataset was created with 7,267 series from 525 subjects, including CH4, CH2, and SAX stacks for each. All test data were retro-gated cine scans from 3T scanners. The median signal-to-noise ratio (SNR) was used to evaluate image quality, with raw SNR measured using the SNRUnit reconstruction, which normalizes noise standard deviation (34). R=4 g-factor maps $g_{R=4}$ were computed for each test case (33). The global median SNR is computed as $median(SNR_{original}/\sqrt{1.0 + (nn \cdot g_{R=4})^2})$ . Here $nn$ is the added noise standard deviation.

Ten target global SNR levels were tested: [0.05, 0.1, 0.2, 0.5, 0.75, 1.0, 1.5, 2.0, 4.0, 8.0], and SD of added noise was computed for each level. As shown in Figure 2, at low SNR, myocardium and blood-pool contrast were lost, and anatomical structures became uninterpretable to human readers (also see Supplement Movie 1).

Four imaging transformer models were trained by scaling up the HRNet backbone: a) IT-27m with block FLG (27 million parameters); b) IT-55m with block FLGFLG; c) IT-109m with block FLGFLGFLGFLG; d) IT-218m with block FLGFLGFLGFLGFLGFLGFLGFLG. Seven baseline models were trained for comparison: ViT3D (25,26) for 27m and 55m, Swin3D (22) for 55m, CNNT (39) for 27m and 55m and convolutional HRNet (28) for 23m and 45m. Their parameters were chosen to match the two smaller IT models. Peak SNR (PSNR, $10 \cdot log_{10}(\frac{MAX^2}{MSE})$) and structural similarity index measure (SSIM) (40) were computed for model outputs. $MSE$ is the mean square difference and $MAX$ is the maximal image value, set to 2048.0.

Two cardiologists (RHD and TAT, >10 years' experience, level III EACVI) reviewed model outputs for 55 randomly selected test cases. GT and model outputs at ten SNR levels were presented together as movies (examples in Supplement Movie 2). Each expert independently picked the lowest SNR offering sufficiently good quality for clinical interpretation. Good quality was defined as no difference in clinical interpretation compared to GT, with adequate contrast, preservation of important anatomical features, and the same subjective assessment of cardiac function.

Cardiac function was quantitatively evaluated by applying a pre-trained, clinically validated cine segmentation model (41), to the GT and model outputs at all SNR levels. This model segmented the left ventricle and myocardium. The model EF was compared to GT values.

The segmentation masks from the GT were used to measure blood-pool and myocardial signal levels, comparing them to the model outputs to assess how well the model recovered the blood-myocardial contrast.

## Results

Figure 3 presents model comparison results. IT-55m was evaluated against other models with similar sizes for SSIM and PSNR at various SNR levels. Both SSIM and PSNR decreased with lower SNR, with IT models outperforming others across all levels. At higher SNR (8.0), all models exhibited excellent SSIM (>0.95), but noticeable differences appeared at lower SNR. For instance, at SNR 0.2, IT-55m had the highest SSIM of 0.8504, outperforming Swin3D-55m (0.6739) and ViT3D-55m (0.7261). Smaller models performed worse; IT-27m had SSIM 0.8276, and ViT3D-27m had 0.6179. Figure 3a shows an example test case for all models and SNR levels. At higher SNR, all models produced good quality and well delineated anatomical details (e.g. the mitral valve). At lower SNR, IT-55m outperformed others, consistent with SSIM and PSNR metrics. Supplement Movie 3 provides zoomed-in videos.

Figure 4 compares the four IT models, showing that performance improved with model size, particularly at low SNR. At SNR 0.2, IT-218m achieved an SSIM of 0.8717, higher than IT-109m (0.8596), IT-55m (0.8504), and IT-27m (0.8276). At SNR 0.1, the SSIM values were 0.7736, 0.7544, 0.7240, and 0.6829, respectively. Figure 4a provides an example for the four models, and Supplement Movie 4 shows corresponding videos. Larger models better preserved anatomical details and tissue contrast. The IT-218m showed loss of details at SNR 0.05, indicating its limit.

Supplement Movies 5, 6, and 7 show IT-218m performance across different cardiac views at various SNR levels (2.0, 0.5, 0.2, 0.1). The model was robust across the SNR range, with significant restoration of image quality, even for fine details like valve motion and papillary muscles, even at low SNR.

Two cardiologists reviewed outputs from 55 test cases, identifying the lowest SNR level for confident clinical interpretation. The median among both experts was that the IT model restored image quality for diagnosis at SNR 0.2 and above. Supplement Movie 8 demonstrates noisy inputs, model outputs, and ground truths at SNR 0.2.

The cine analysis model was used to segment the LV and myocardium, and compute EF. Figure 5a shows Bland-Altman plots of EF for IT-218m and ground truth at all SNR levels. The mean deviation at SNR 8.0 was -0.08335%, and at SNR 0.2, increased to 0.7975%. The 90% confidence range (CR) increased with lower SNR. For context, the cine analysis model has a test-retest reproducibility of 4.6% for EF (41), which was higher or on par with the CR up to SNR 0.2. Figure 5b-e gives an example to illustrate the analysis. Supplement Movie 9 shares the corresponding movies.

Smaller models like IT-27m had inferior image quality and higher errors in EF estimates. Figure E1 [supplement] shows the EF Bland-Altman plots for IT-27m. At SNR 8.0, IT-27m showed a mean deviation of -0.09215%, comparable to IT-218m. At SNR 0.2, the error increased to 4.463%, much higher than IT-218m. The same trend was found for CR. When SNR was lower than 0.5, the IT-27m had higher scatter of errors than the IT-218m. Figure E1c [supplement] compared the 90% CRs of IT-27m against the IT-218m.

Figure 6a and 6b give the Bland-Altman plots of blood and myocardial model signals against the GT. The mean signals for blood and myocardium were 105.7 and 37.0. The deviation in blood signal was less than 0.08% at SNR 8.0 and 1.26% at SNR 0.2. For myocardium, they were 0.09% and 1.55%. As a result, the blood-myocardial contrast was restored, as shown in Figure 6e and Supplement Movie 10.

The model's robustness against varying acceleration factors is demonstrated in Figure E2 (supplement) with g-factor maps from R=2 to R=5. At R=5, the g-factor noise amplification was severe. The model effectively denoised spatially amplified noises across a range of starting SNRs, with performance starting to slip only at SNR ~0.2. Supplement Movie 11 and 12 show videos before and after model application.

## Discussion

Recovering signals from noisy data is crucial for high-performance imaging. A key finding is that large IT models can recover image quality sufficient for clinical reading, even at very low SNRs. When noise is five times higher than the signal, the blood-myocardial contrast is lost, and fine anatomical structures are noninterpretable. However, the model still recovers signal faithfully, producing EF estimates, showcasing the power of large IT models.

The core innovation of imaging transformers lies in decoupling full attention into spatial local, spatial global, and frame modules. This reduces quadratic complexity, offers versatility and flexibility, and enables focused attention on intra- and inter-frame dimensions. A module–cell–block architecture allows for model scaling. Using a HRNet backbone, IT models were trained on a large MR denoising dataset, outperforming baselines using ViT, Swin, and CNNs, with the performance gap widening at lower SNRs. Scaling IT models from 27 million to 218 million parameters improved performance across all SNR levels.

Other studies in language modelling have attempted to address quadratic complexity of the classical scale-dot-product attention (42) through sparse (43–45) or low-rank approximations (20,46,47). The IO-aware Flash-attention (48,49) speeds up attention without approximation using kernel fusion, re-computation, and multi-threading. For images, MaxViT (20) decouples full attention of a 2D image into local and remote components, which is the same spatial splitting as our approach. However, we use convolutions to maintain locality bias, handle varying matrix sizes, and reduce number of parameters, offering an effective low-rank approximation by limiting the attention matrix computation.

With a large training dataset, we showed that IT models can scale for better performance. Larger models achieved higher SSIM and PSNR across SNR levels, and IT-218m produced more accurate EF estimates. The error for the 27m model increased 5.6× at SNR 0.2, highlighting the importance of scaling for accurate biomarker estimation. Unlike other studies focused on SSIM or loss metrics, our work specifically evaluated clinical relevance. To our best knowledge, no other studies have verified biomarker estimation can benefit from model scaling. One other study scaled the U-net for MR reconstruction (50) to measure SSIM. The language model scaling was assessed for cardiac MR tasks (49) to compare the cross-entropy loss.

This study has limitations. We focused on introducing the IT model and evaluating performance across a range of SNR levels with available ground-truth data and did not target more generalization cases. While we assessed EF here, other biomarkers are also relevant. Clinical applicability could be tested in more imaging sequences and conditions. Lastly, although we demonstrated scaling up to 218 million parameters, further scaling requires advancements in model software and training methods. More computing is needed for larger datasets. All of these are topics for future research.

**Availability of data and material**

The authors thank the Intramural Research Program of the National Heart, Lung, and Blood Institute for the data obtained from the NIH Open-Source Cardiac MRI Raw Data Repository.

**Funding**

None

**Authors' contributions**

--

## References

1. Chung H, Lee ES, Ye JC. MR Image Denoising and Super-Resolution Using Regularized Reverse Diffusion. IEEE Trans Med Imaging. 2023;42(4):922–934. doi: 10.1109/TMI.2022.3220681.
2. Tian C, Fei L, Zheng W, Xu Y, Zuo W, Lin C-W. Deep Learning on Image Denoising: An overview. Neural Networks. 2020;(131):251–275. doi: https://doi.org/10.1016/j.neunet.2020.07.025.
3. Dou Q, Wang Z, Feng X, Campbell-Washburn AE, Mugler JP, Meyer CH. MRI denoising with a non-blind deep complex-valued convolutional neural network. NMR Biomed. John Wiley and Sons Ltd; 2024; doi: 10.1002/nbm.5291.
4. Shou Q, Zhao C, Shao X, et al. Transformer-based deep learning denoising of single and multi-delay 3D arterial spin labeling. Magn Reson Med. John Wiley and Sons Inc; 2024;91(2):803–818. doi: 10.1002/mrm.29887.
5. Dziadosz M, Rizzo R, Kyathanahally SP, Kreis R. Denoising single MR spectra by deep learning: Miracle or mirage? Magn Reson Med. John Wiley and Sons Inc; 2023;90(5):1749–1761. doi: 10.1002/mrm.29762.
6. Kurmi Y, Viswanathan M, Zu Z. A Denoising Convolutional Autoencoder for SNR Enhancement in Chemical Exchange Saturation Transfer imaging: (DCAE-CEST). 2024. doi: 10.1101/2024.06.07.597818.
7. Kang B, Lee W, Seo H, Heo HY, Park HW. Self-supervised learning for denoising of multidimensional MRI data. Magn Reson Med. John Wiley and Sons Inc; 2024;92(5):1980–1994. doi: 10.1002/mrm.30197.
8. Lebel RM, Healthcare GE. Performance characterization of a novel deep learning-based MR image reconstruction pipeline. .
9. Salehi A, Mach M, Najac C, et al. Denoising low-field MR images with a deep learning algorithm based on simulated data from easily accessible open-source software. Journal of Magnetic Resonance. Academic Press Inc.; 2025;370. doi: 10.1016/j.jmr.2024.107812.
10. Pfaff L, Hossbach J, Preuhs E, et al. Self-supervised MRI denoising: leveraging Stein's unbiased risk estimator and spatially resolved noise maps. Sci Rep. Nature Research; 2023;13(1). doi: 10.1038/s41598-023-49023-2.
11. Kidoh M, Shinoda K, Kitajima M, et al. Deep learning based noise reduction for brain MR imaging: Tests on phantoms and healthy volunteers. Magnetic Resonance in Medical Sciences. Japanese Society for Magnetic Resonance in Medicine; 2020;19(3):195–206. doi: 10.2463/mrms.mp.2019-0018.
12. Koonjoo N, Zhu B, Bagnall GC, Bhutto D, Rosen MS. Boosting the signal-to-noise of low-field MRI with deep learning image reconstruction. Sci Rep. Nature Research; 2021;11(1). doi: 10.1038/s41598-021-87482-7.
13. Bash S, Wang L, Airriess C, et al. Deep learning enables 60% accelerated volumetric brain MRI while preserving quantitative performance: A prospective, multicenter, multireader trial. American Journal of Neuroradiology. American Society of Neuroradiology; 2021. p. 2130–2137. doi: 10.3174/ajnr.A7358.
14. Xiang T, Yurt M, Syed AB, Setsompop K, Chaudhari A. DDM 2 : SELF-SUPERVISED DIFFUSION MRI DENOIS-ING WITH GENERATIVE DIFFUSION MODELS. ICLR. 2023. https://github.com/StanfordMIMI/DDM2.

15. He K, Zhang X, Ren S, Sun J. Deep Residual Learning for Image Recognition. 2015; http://arxiv.org/abs/1512.03385.
16. Li J, Chen J, Tang Y, Wang C, Landman BA, Zhou SK. Transforming medical imaging with Transformers? A comparative review of key properties, current progresses, and future perspectives. 2022; http://arxiv.org/abs/2206.01136.
17. Shamshad F, Khan S, Zamir SW, et al. Transformers in Medical Imaging: A Survey. 2022; http://arxiv.org/abs/2201.09873.
18. Waqas Zamir S, Arora A, Khan S, Hayat M, Shahbaz Khan F, Yang M-H. Restormer: Efficient Transformer for High-Resolution Image Restoration. . https://github.com/swz30/Restormer.
19. Wang Z, Cun X, Bao J, Zhou W, Liu J, Li H. Uformer: A General U-Shaped Transformer for Image Restoration. 2021; http://arxiv.org/abs/2106.03106.
20. Tu Z, Talebi H, Zhang H, et al. MaxViT: Multi-Axis Vision Transformer. 2022; http://arxiv.org/abs/2204.01697.
21. Liu Z, Lin Y, Cao Y, et al. Swin Transformer: Hierarchical Vision Transformer using Shifted Windows. . https://github.
22. Liang J, Cao J, Sun G, Zhang K, Van Gool L, Timofte R. SwinIR: Image Restoration Using Swin Transformer. 2021; http://arxiv.org/abs/2108.10257.
23. Luo W, Li Y, Urtasun R, Zemel R. Understanding the Effective Receptive Field in Deep Convolutional Neural Networks. 2017; http://arxiv.org/abs/1701.04128.
24. Vaswani A, Brain G, Shazeer N, et al. Attention Is All You Need. .
25. Dosovitskiy A, Beyer L, Kolesnikov A, et al. An Image is Worth 16x16 Words: Transformers for Image Recognition at Scale. 2020; http://arxiv.org/abs/2010.11929.
26. Ali AM, Benjdira B, Koubaa A, El-Shafai W, Khan Z, Boulila W. Vision Transformers in Image Restoration: A Survey. Sensors. MDPI; 2023;23(5). doi: 10.3390/s23052385.
27. Ronneberger O, Fischer P, Brox T. U-Net: Convolutional Networks for Biomedical Image Segmentation. 2015; http://arxiv.org/abs/1505.04597.
28. Wang J, Sun K, Cheng T, et al. Deep High-Resolution Representation Learning for Visual Recognition. 2019; http://arxiv.org/abs/1908.07919.
29. Shaw P, Uszkoreit J, Vaswani A. Self-Attention with Relative Position Representations. 2018; http://arxiv.org/abs/1803.02155.
30. Ba JL, Kiros JR, Hinton GE. Layer Normalization. 2016; http://arxiv.org/abs/1607.06450.
31. He K, Zhang X, Ren S, Sun J. Delving Deep into Rectifiers: Surpassing Human-Level Performance on ImageNet Classification. 2015; http://arxiv.org/abs/1502.01852.
32. Hinton GE, Srivastava N, Krizhevsky A, Sutskever I, Salakhutdinov RR. Improving neural networks by preventing co-adaptation of feature detectors. 2012; http://arxiv.org/abs/1207.0580.
33. Xue H, Hooper SM, Pierce I, et al. SNRAware: Improved Deep Learning MRI Denoising with SNR Unit Training and G-factor Map Augmentation. .
34. Kellman P, McVeigh ER. Image reconstruction in SNR units: A general method for SNR measurement. Magn Reson Med. 2005;54(6):1439–1447. doi: 10.1002/mrm.20713.
35. Pruessmann KP, Weiger M, Scheidegger MB, Boesiger P. SENSE: Sensitivity encoding for fast MRI. Magn Reson Med. 1999;42(5):952–962. doi: 10.1002/(SICI)1522-2594(199911)42:5<952::AID-MRM16>3.0.CO;2-S.

36. Liu H, Li Z, Hall D, Liang P, Ma T. SOPHIA: A SCALABLE STOCHASTIC SECOND-ORDER OPTIMIZER FOR LANGUAGE MODEL PRE-TRAINING. .
37. Smith LN. A disciplined approach to neural network hyper-parameters: Part 1 -- learning rate, batch size, momentum, and weight decay. 2018; http://arxiv.org/abs/1803.09820.
38. Paszke A, Gross S, Massa F, et al. PyTorch: An Imperative Style, High-Performance Deep Learning Library. 2019.
39. Rehman A, Zhovmer A, Sato R, et al. Convolutional neural network transformer (CNNT) for fluorescence microscopy image denoising with improved generalization and fast adaptation. Sci Rep. Nature Research; 2024;14(1). doi: 10.1038/s41598-024-68918-2.
40. Wang Z, Bovik AC, Sheikh HR, Simoncelli EP. Image quality assessment: From error visibility to structural similarity. IEEE Transactions on Image Processing. 2004;13(4):600–612. doi: 10.1109/TIP.2003.819861.
41. Davies RH, Augusto JB, Bhuva A, et al. Precision measurement of cardiac structure and function in cardiovascular magnetic resonance using machine learning. Journal of Cardiovascular Magnetic Resonance. BioMed Central; 2022;24(1):1–11. doi: 10.1186/s12968-022-00846-4.
42. Vaswani A, Shazeer N, Parmar N, et al. Attention is all you need. Adv Neural Inf Process Syst. 2017;2017-Decem(Nips):5999–6009.
43. Child R, Gray S, Radford A, Sutskever I. Generating Long Sequences with Sparse Transformers. 2019; http://arxiv.org/abs/1904.10509.
44. Kitaev N, Kaiser Ł, Levskaya A. Reformer: The Efficient Transformer. 2020; http://arxiv.org/abs/2001.04451.
45. Ren H, Dai H, Dai Z, et al. Combiner: Full Attention Transformer with Sparse Computation Cost. 2021; http://arxiv.org/abs/2107.05768.
46. Chen B, Dao T, Winsor E, Song Z, Rudra A, Ré C. Scatterbrain: Unifying Sparse and Low-rank Attention Approximation. 2021; http://arxiv.org/abs/2110.15343.
47. Choromanski K, Likhosherstov V, Dohan D, et al. Rethinking Attention with Performers. 2020; http://arxiv.org/abs/2009.14794.
48. Dao T. FlashAttention-2: Faster Attention with Better Parallelism and Work Partitioning. 2023; http://arxiv.org/abs/2307.08691.
49. Dao T, Fu DY, Ermon S, Rudra A, Ré C. FlashAttention: Fast and Memory-Efficient Exact Attention with IO-Awareness. 2022; http://arxiv.org/abs/2205.14135.
50. Klug T, Heckel R. Scaling Laws For Deep Learning Based Image Reconstruction. 2022; http://arxiv.org/abs/2209.13435.
51. Hooper S, Xue H. A Study on Context Length and Efficient Transformers for Biomedical Image Analysis. Proc Mach Learn Res. 2024.

**Figure 1**. Imaging attention modules and model design. (a) Proposed model recovered signal faithfully from very low SNR input. (b) We proposed the 5D tensor [B, C, F, H, W] as a key data format to represent imaging data, for 2D, 2D+T or 3D acquisition. Full attention across three dimensions can incur very high computing costs and may not be optimal for not exploiting the locality of information. Instead, intra- and inter-frame attention can be separately computed. For the spatial (along H, W) attention, a target patch (marked in yellow) can take in information from both close neighbors (marked in green) and remote patches (marked in pink). We showed this approximation method mitigates quadratic complexity of full attention, brings back inductive bias and is very effective in MR denoising. (c) Three attention modules are illustrated. Spatial attention splits an image to multiple windows. Each window is further divided into patches. Local attention computes the attention matrix with all patches in one window. Global attention computes the attention over the corresponding patches from all windows. For example, one attention is computed over all blue patches and another attention is computed over all red ones. Inter-frame correlation is captured with frame attention. (d) Following the classical transformer design, a cell holds an attention module, layer normalization and the mixer, with the skip connections. A cell can hold either G, L or T attention, to aggregate information from different regions. A block is a container for many cells. Model can be scaled up in size by inserting more blocks or inserting more cells into every block. (e) A HRNet backbone was instantiated to hold five blocks at two resolution levels. The number of channels after downsample was doubled. Every block takes in a 5D tensor and produces another 5D tensor as the output. By doing so, all components can be concatenated. For the denoising task, we also add pre- and post-convolution to convert input 3 channels (real, imag and g-factor) to C=64 internal channels. Output conv will produce 2 channels for complex data.

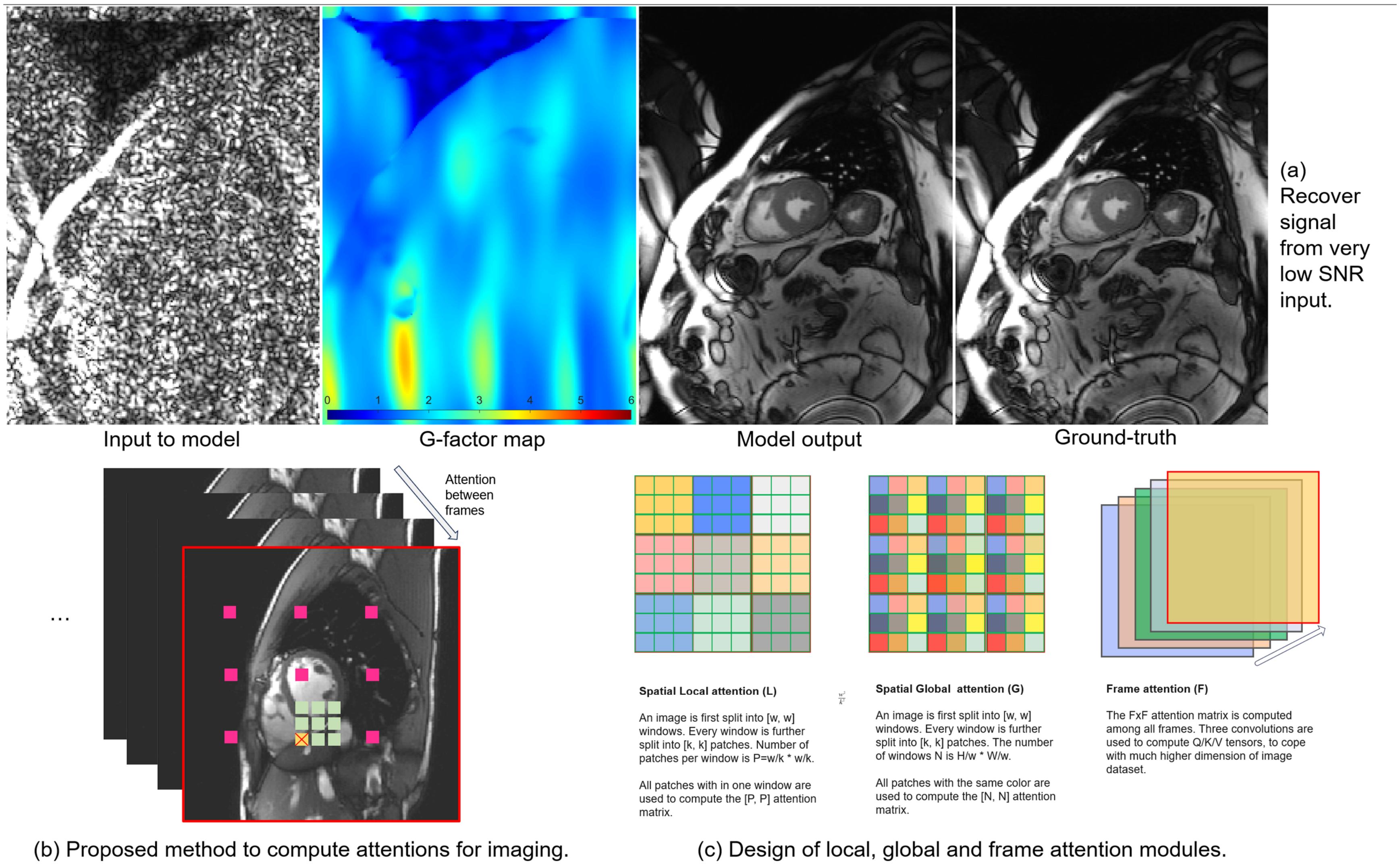
0 1 2 3 4 5 6
Input to model
G-factor map
Model output
Ground-truth
(a) Recover signal from very low SNR input.
...
Attention between frames
(b) Proposed method to compute attentions for imaging.
Spatial Local attention (L)
An image is first split into [w, w] windows. Every window is further split into [k, k] patches. Number of patches per window is P=w/k * w/k.
All patches with in one window are used to compute the [P, P] attention matrix.
Spatial Global attention (G)
An image is first split into [w, w] windows. Every window is further split into [k, k] patches. The number of windows N is H/w * W/w.
All patches with the same color are used to compute the [N, N] attention matrix.
Frame attention (F)
The FxF attention matrix is computed among all frames. Three convolutions are used to compute Q/K/V tensors, to cope with much higher dimension of image dataset.
(c) Design of local, global and frame attention modules.

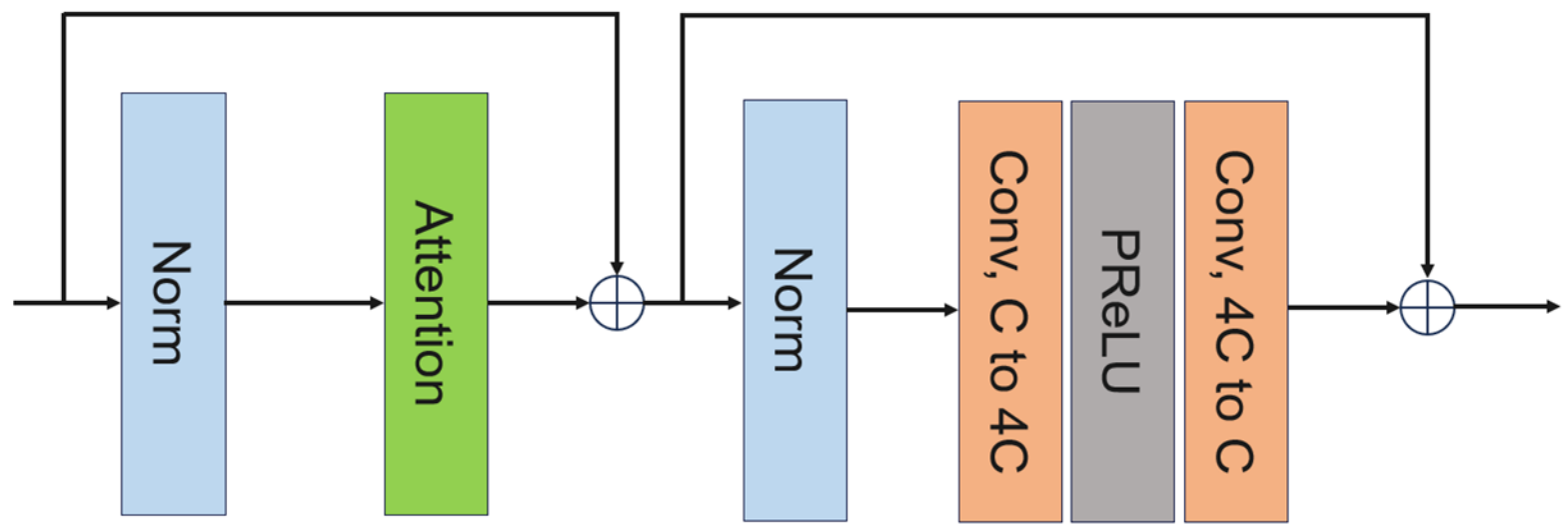

Cell: a cell contains one attention module. It follows the classical design in transformer. The layer normalization was used here. The mixer was implemented as two convolution layers separated by a PReLU nonlinear activation. The attention can be one of F, L or G module. A dropout was added to the cell output to provide regularization.

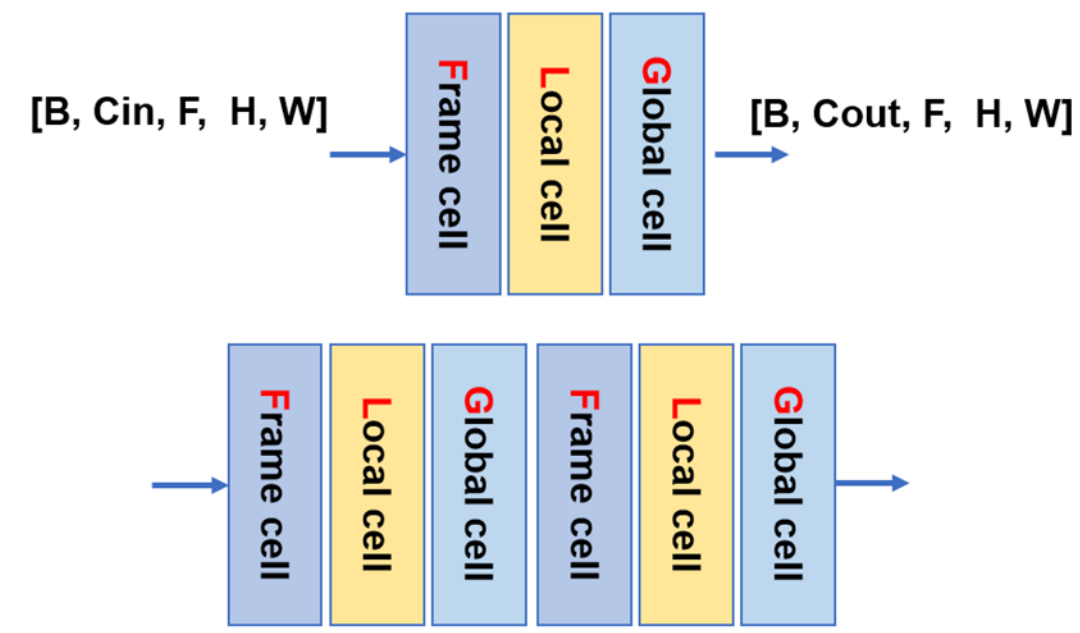

Block: a block contains many attention cells. Its architecture is defined by the building subscription, e.g. FLG means a block consists of three modules. The first is a frame cell. The second and third are spatial local and global attention cells. This modularized design makes scale up the block and model convenient. The block can be doubled to hold "FLGFLG" six attention modules. It also allows to control emphasis of spatial and frame attention; for example, a "FLFGFLFG" block can put more focus on the inter-frame attention.

(d) A modularized design was proposed to define cell and block as two levels of components to build up models. A cell contains one of the attention modules, with normalization, residual connection and mixer. A block contains any number of cells and can be scaled up by inserting more cells.

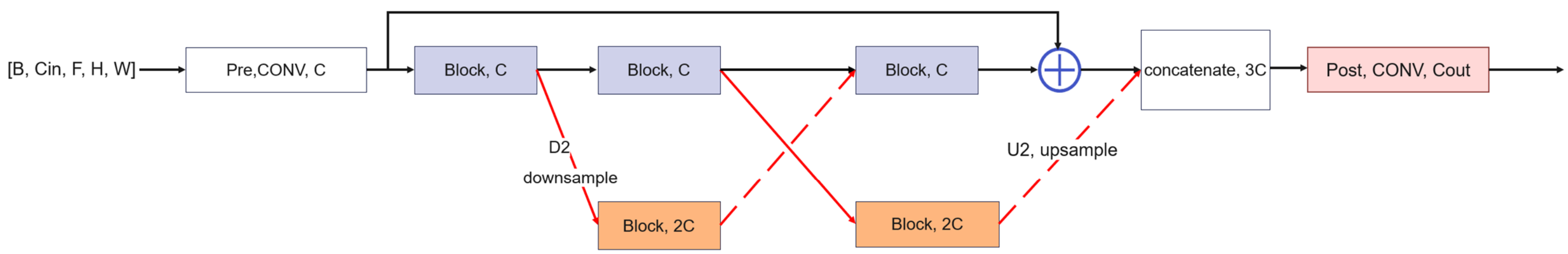

(e) An instantiation of HRNet backbone consists of 5 blocks. The model processes 5D tensors as a general data format for imaging. The tensor sizes were kept unchanged inside blocks and adjusted with downsample and upsample layers. Together with pre and post conv layers, this model was trained for MR denoising.

**Figure 2**. A test data with N=10 levels of signal-noise-ratio. (a) The GT data has a high SNR of 23.6. By sampling noises and amplifying them with g-factor maps, a set of 10 input series with SNR from 8.0 to 0.05 was created test the model. At low SNR end, e.g. SNR<1, the anatomical features are lost to the elevated noise and blood-myocardial contrast decreases. (b) The blood-myocardial contrast-to-noise ratio was plotted against the input SNR. The elevated noise reduces the CNR asymptotically towards zero and makes the image uninterpretable.

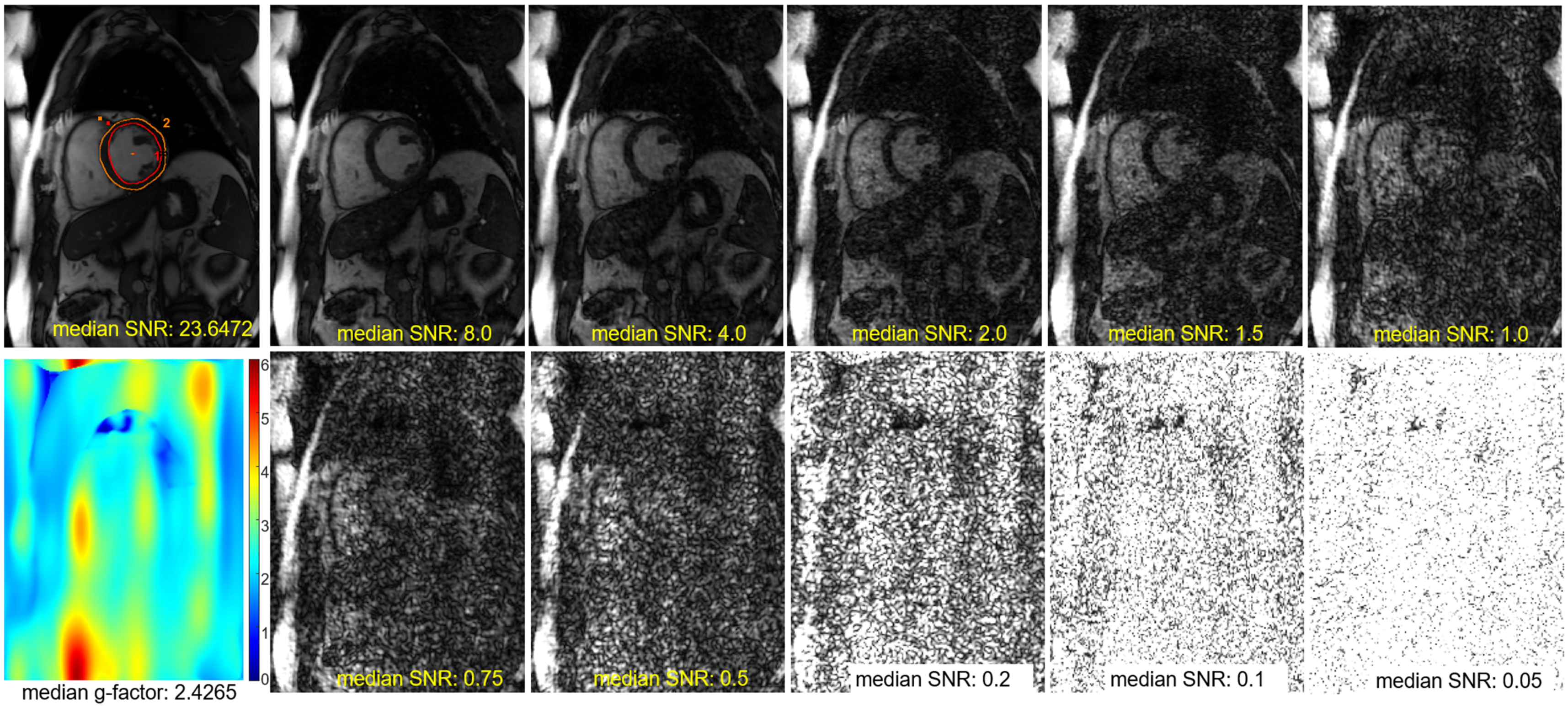

(a) Generated input for model evaluation with different global median signal-to-noise ratios.

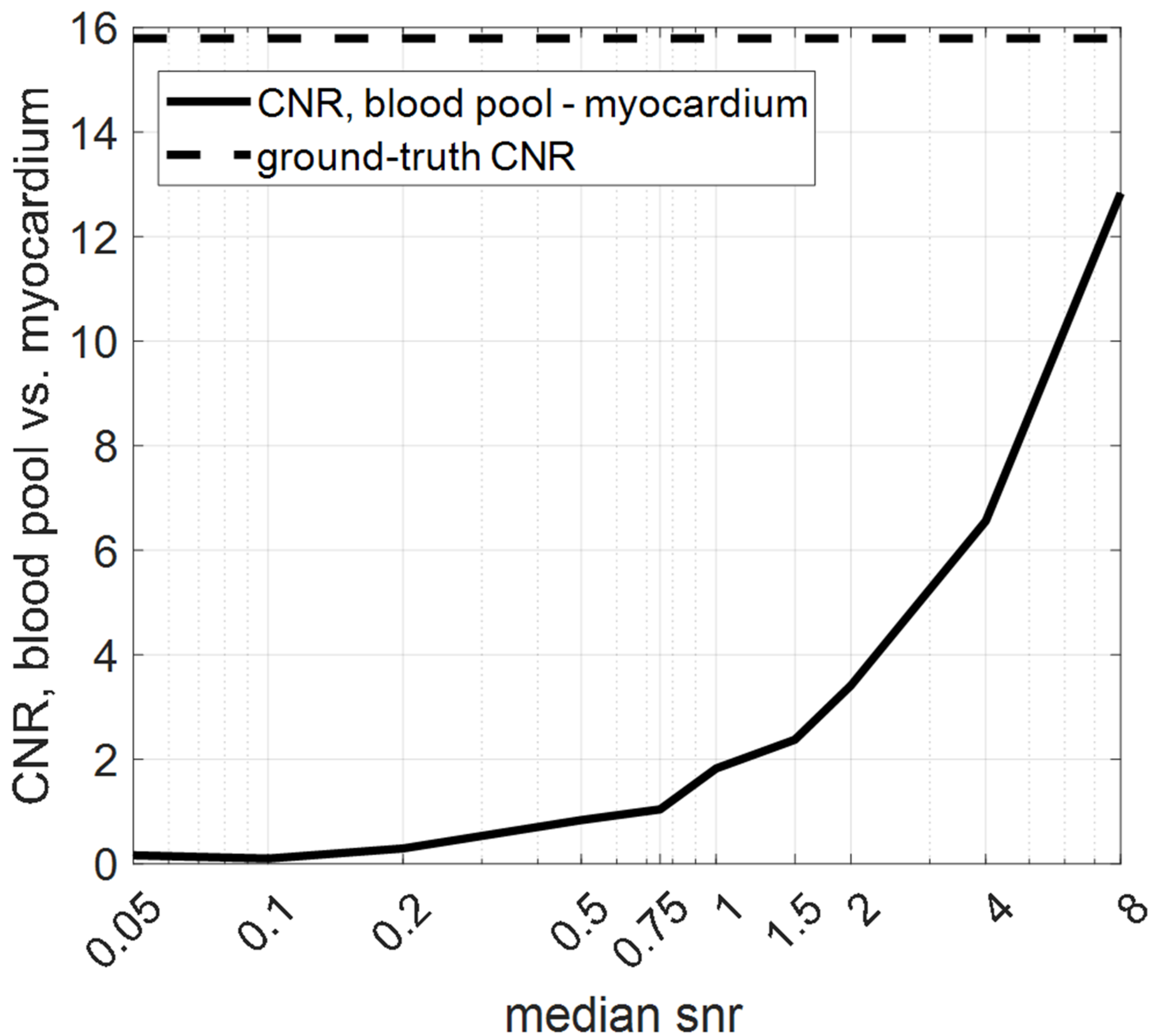

(b) Reducing the SNR leads to loss of contrast. As an example, the CNR between LV blood pool and myocardium was measured and plotted against the median SNR.

**Figure 3**. Comparison to other models (also in Supplement Movie 3). (a) The IT-55m model was compared to other tested models with matching number of parameters. The top row is the input from low SNR of 0.05 to high 8.0. All models worked well at higher SNR end, but imaging transformers showed noticeable lead in overall image quality, especially the recovery of anatomical details when SNR is 0.2 or lower. At SNR 0.05, outputs of all models are non-diagnostic, showing the limit of model and space for further improvement. (b) The test set was processed with all models. SSIM and PSNR were computed over all SNR levels and plotted. The IT-55m model outperformed others consistently across the wide SNR range. The performance gap was more substantial at the low SNR.

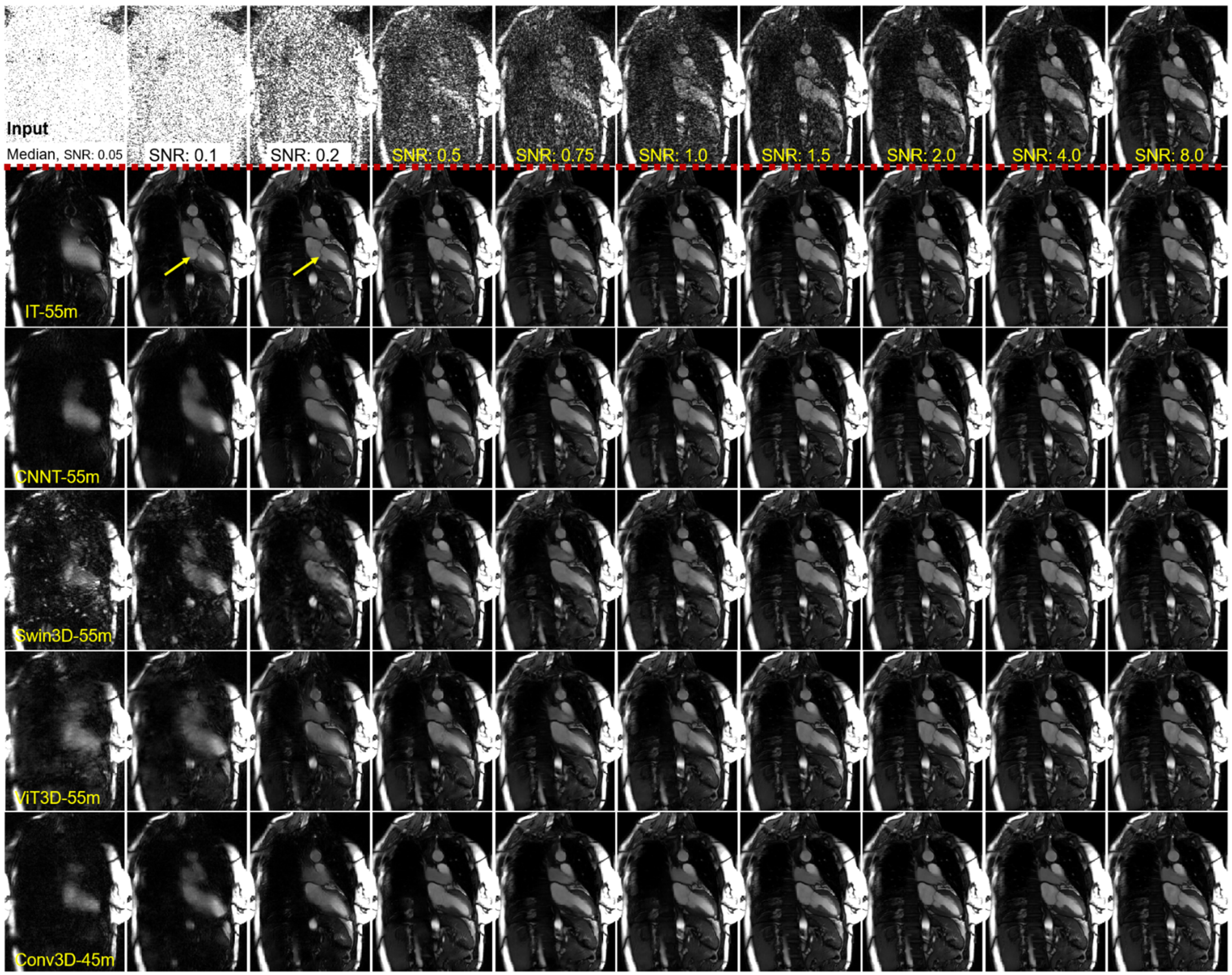

(a) Imaging transformers are compared with other tested models. The first row is the input to models and other rows are inference results.

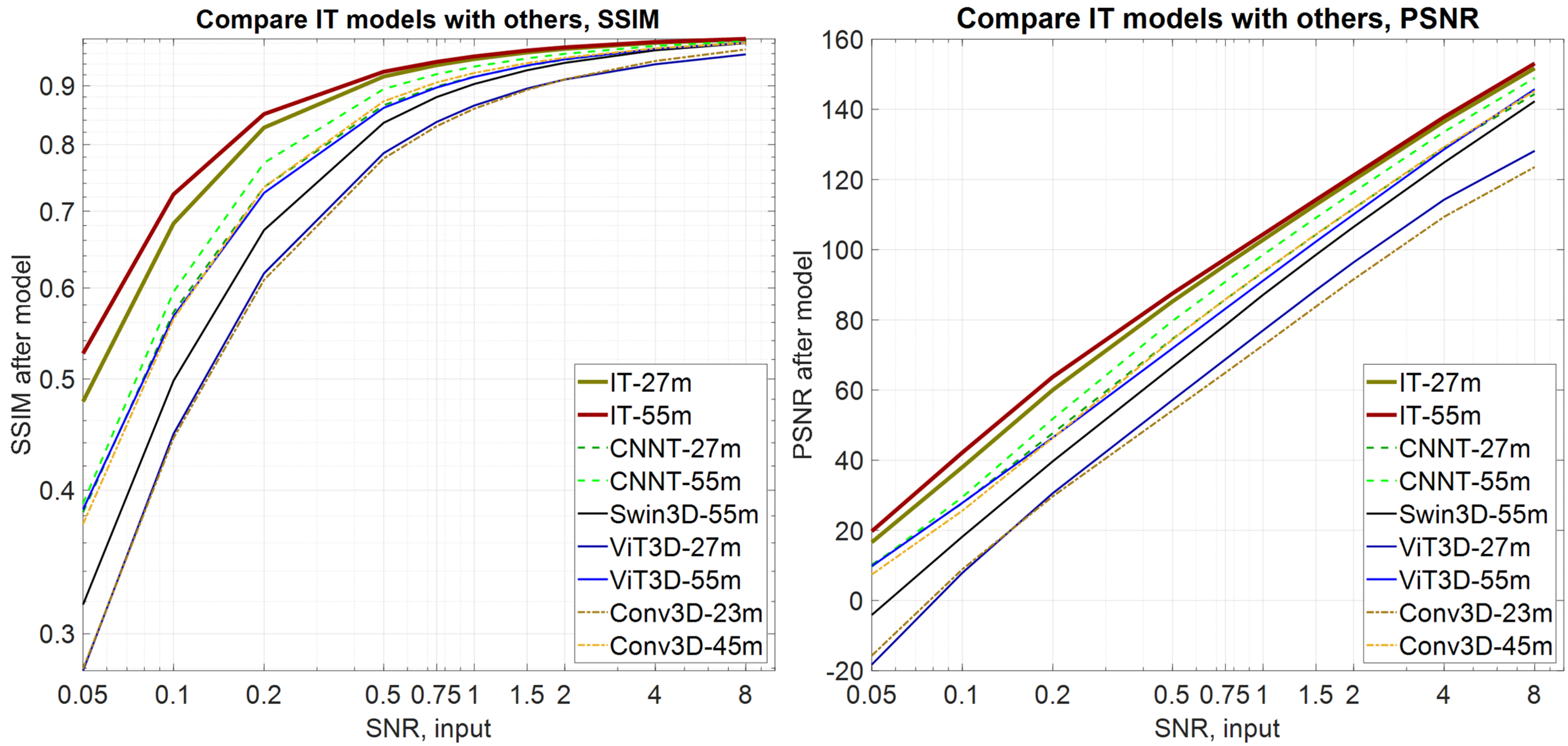

(b) Plots of median SNR against SSIM and PSNR show imaging transformers offer the best performance.

**Figure 4**. Comparison of four imaging transformer models of different sizes (also in Supplement Movie 4). (a) Four IT models were applied to a test example. The bigger model had the best image quality at the low SNR. (b) The curves show that scaling up the model size improves the performance.

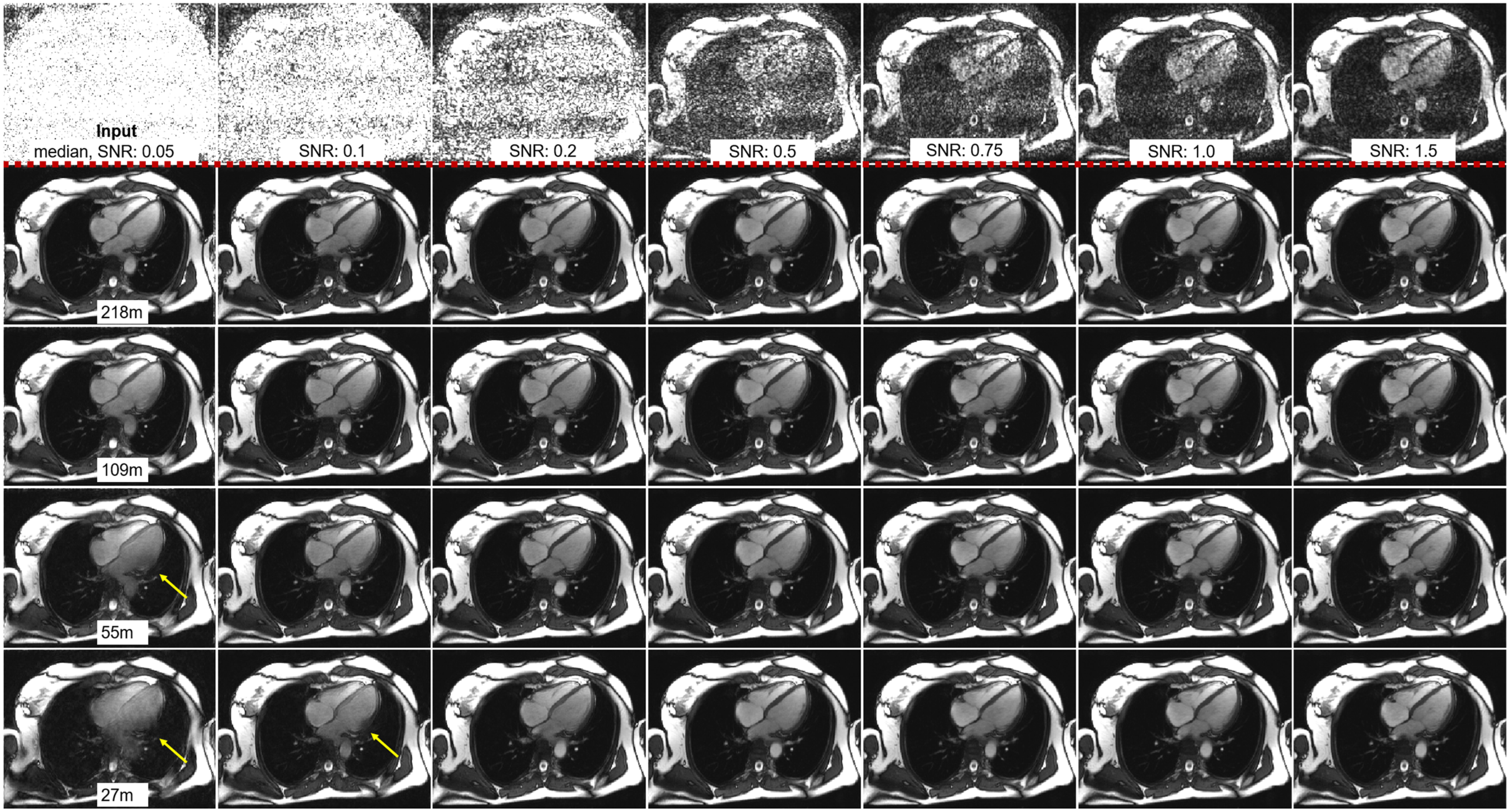

(a) Comparison of four imaging transformer models with different sizes.

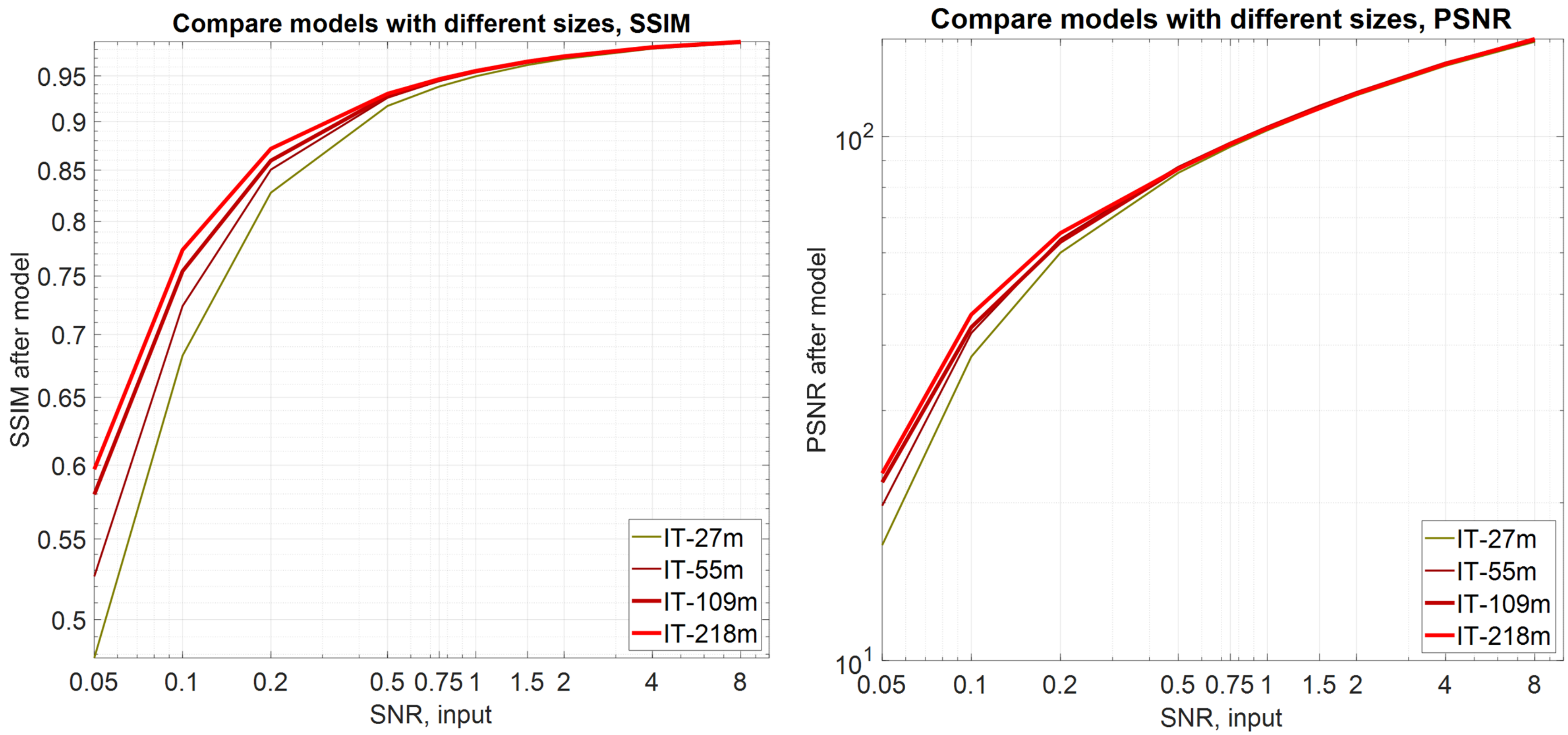

(b) Plots of SSIM and PSNR to compare different model sizes.

**Figure 5**. The ejection fraction measured on IT-218m outputs were compared to ground-truth values. (a) The Bland-Altman plots of EF, GT vs. model over all SNR levels, show that model measurement is correct. The 90% CR increased with lower SNR levels. (b-e) An example of EF measurement is shown here. The cine analysis model was applied to the short-axis stacks of the GT and model outputs, to segment the myocardium and blood pool. The image quality was greatly improved after model, allowing the cine analysis to work well on the model outputs.

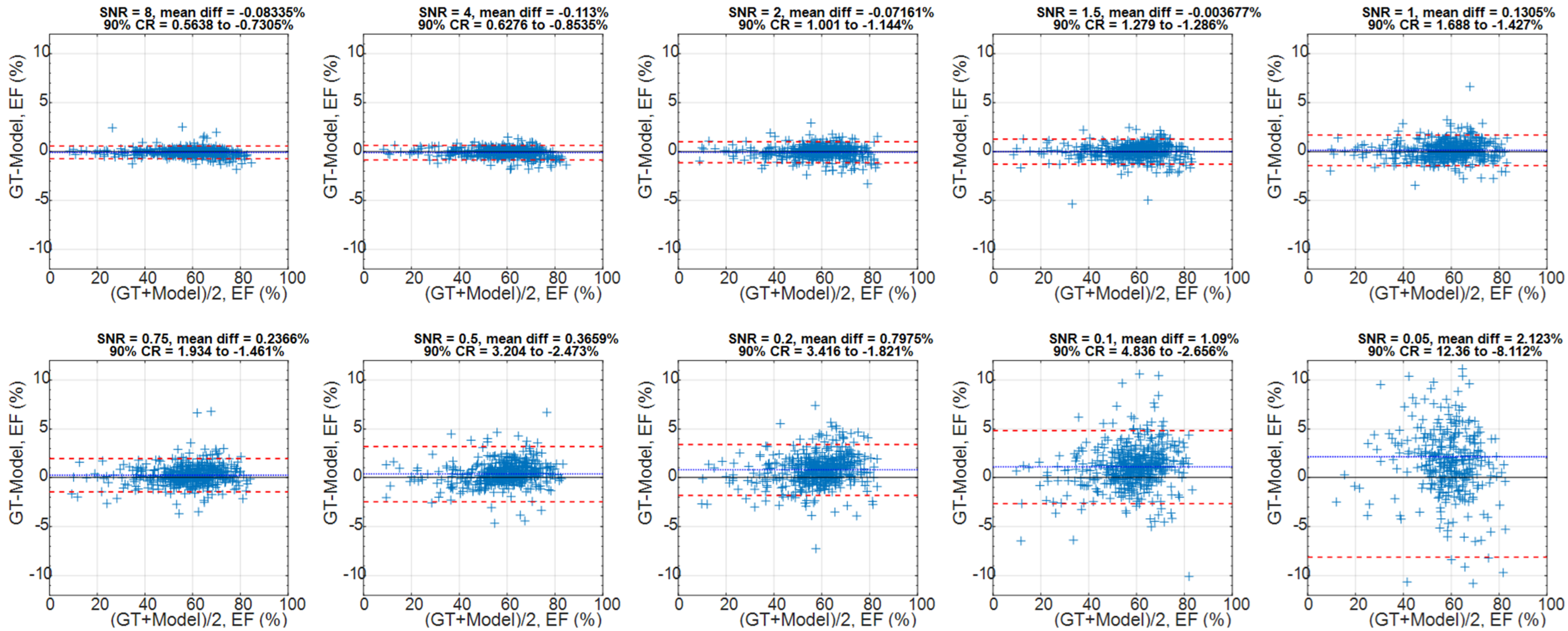

(a) Bland-Altman plots of EF over different SNR levels.

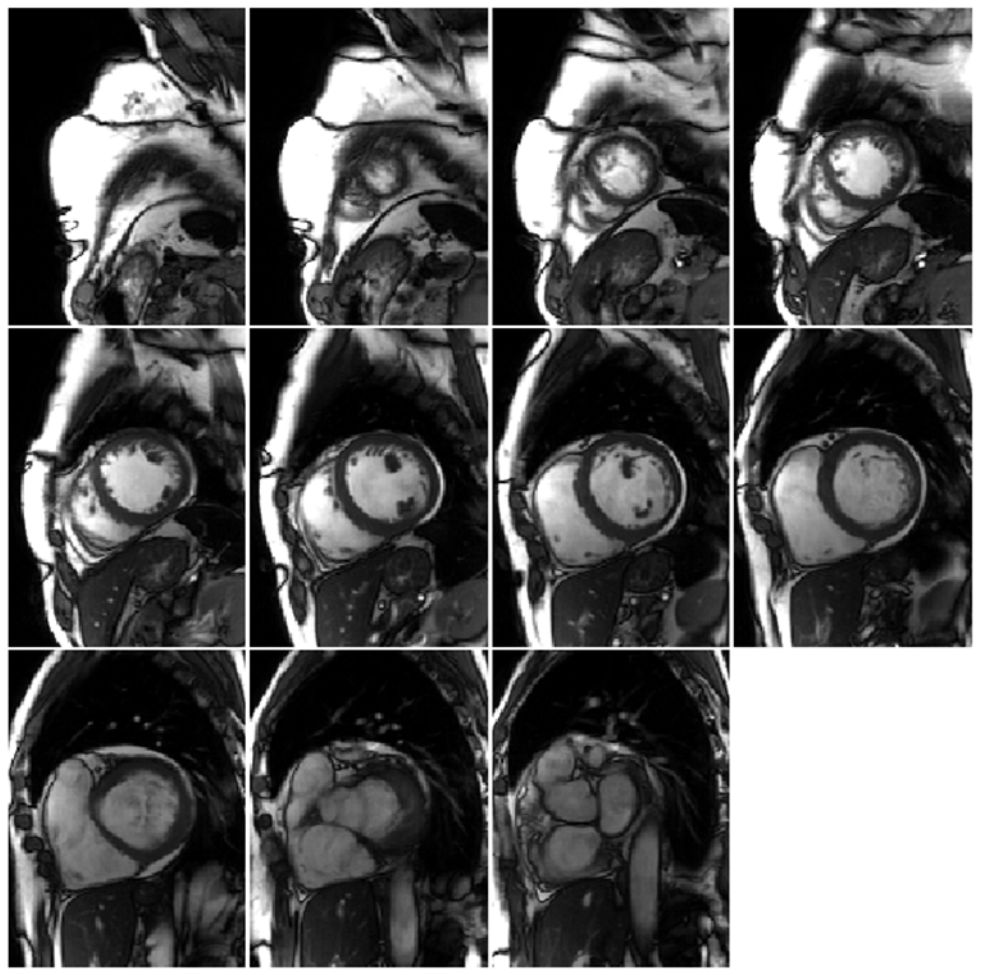

(b) Ground-truth of a short-axis cine stack.

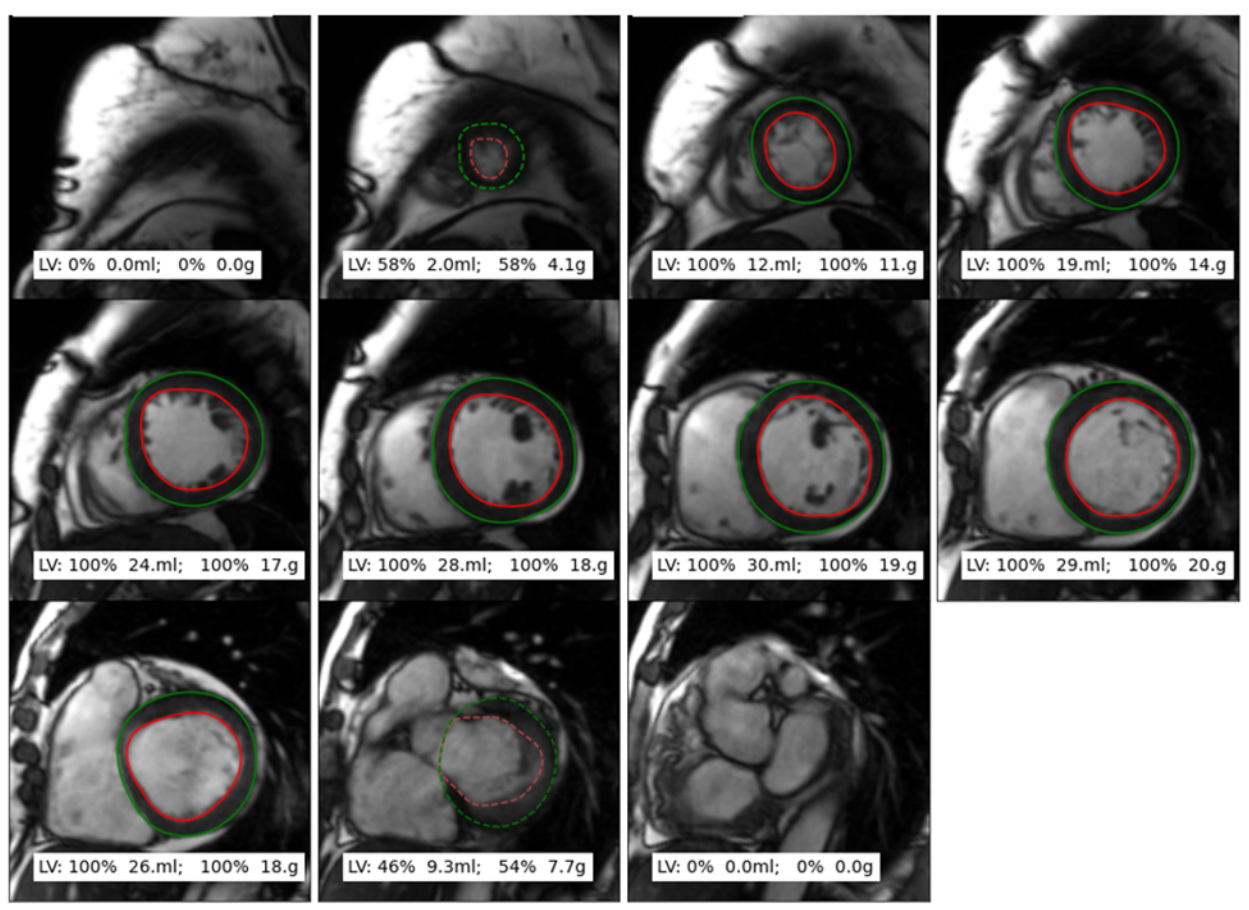

(c) Myocardial segmentation.

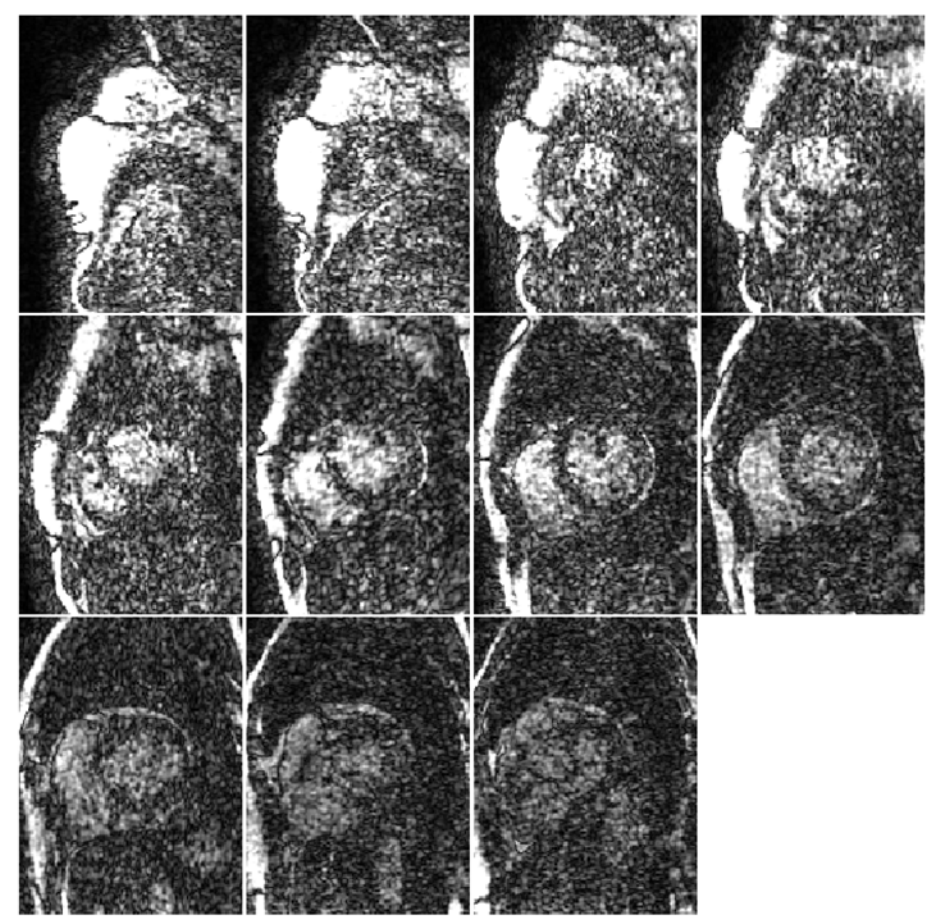

(d) Input to model inference.

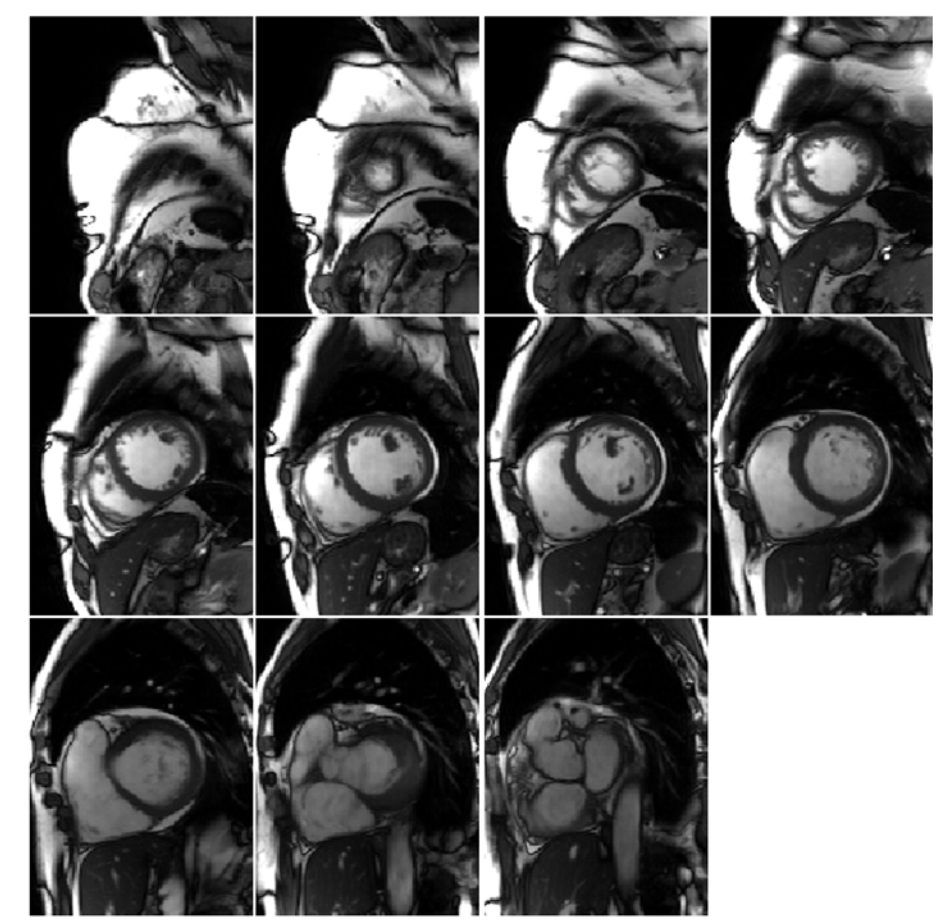

(e) Output after model inference.

**Figure 6.** The blood pool and myocardial signal levels were measured with segmentation masks generated on the GT data. (a-b) The Bland-Altman plots for the blood signal and the myocardium. (c) An example of segmentation mask overlaid on the image. (d) For the starting SNR 0.1, the CNR was reduced from 113.2 to 5.72. After model, the CNR level was restored to be 112.3. (e) The CNR was measured for all SNR levels. The model was able to restore the contrast until ~SNR 0.2. With even lower SNR, the resulting CNR started to deviate from the ground-truth, but was still much higher than starting values.

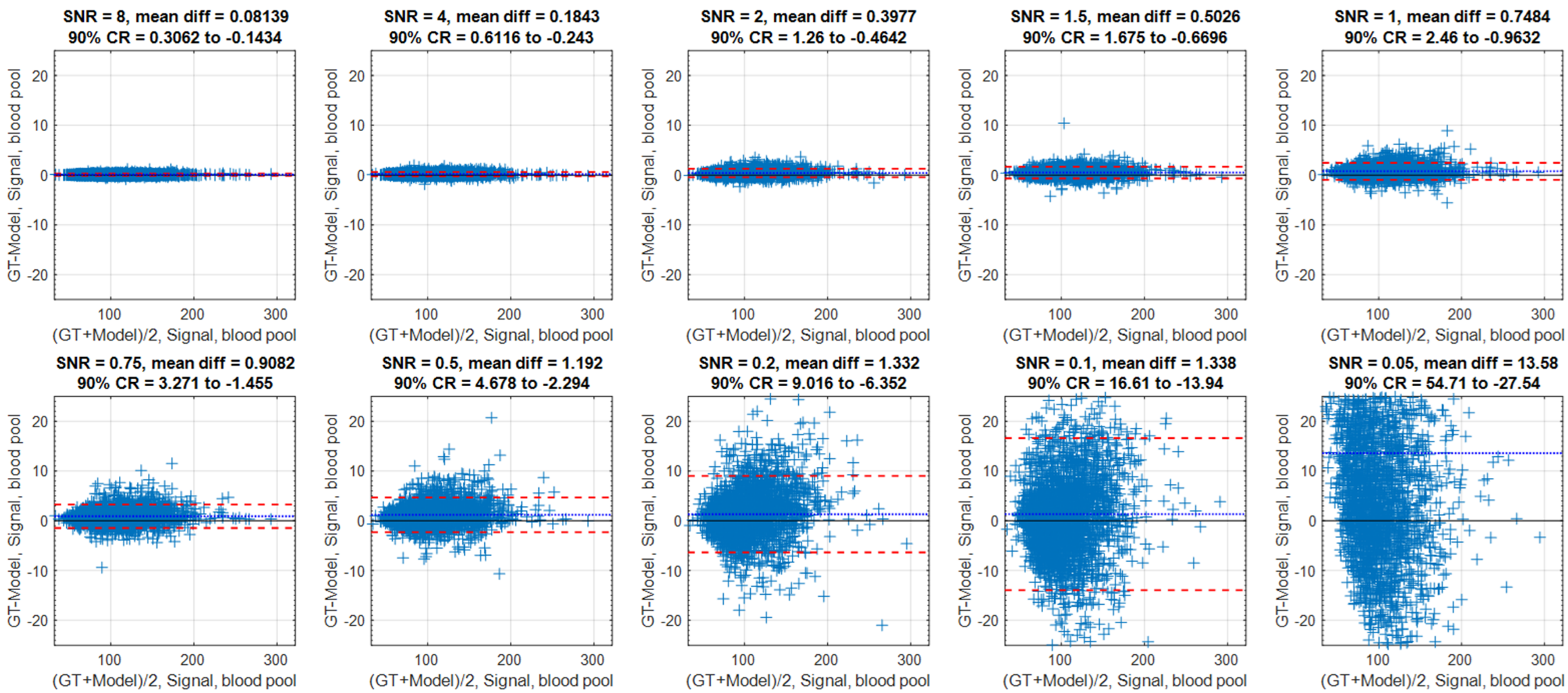

(a) Bland-Altman plots of blood pool signal levels between ground-truth and model outputs.

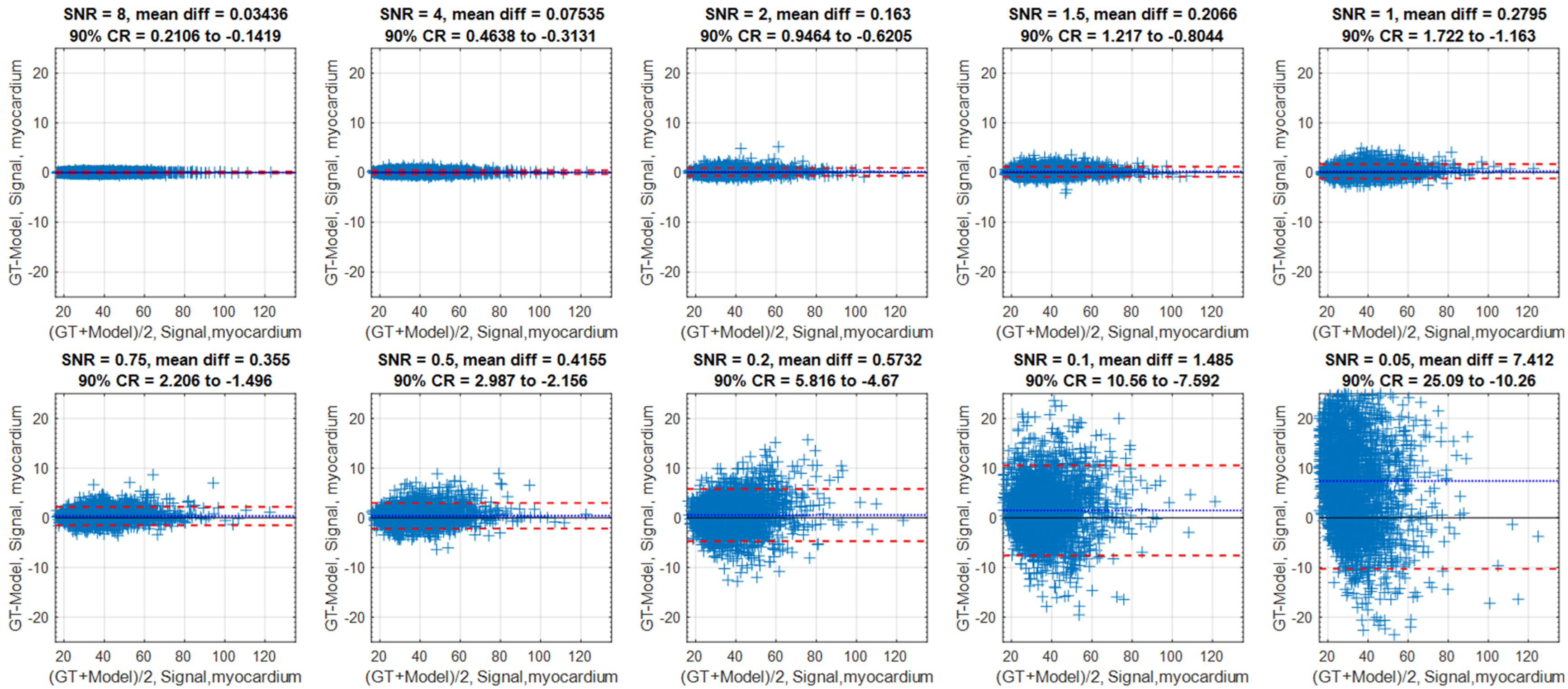

(b) Bland-Altman plots of myocardium signal levels between ground-truth and model outputs.

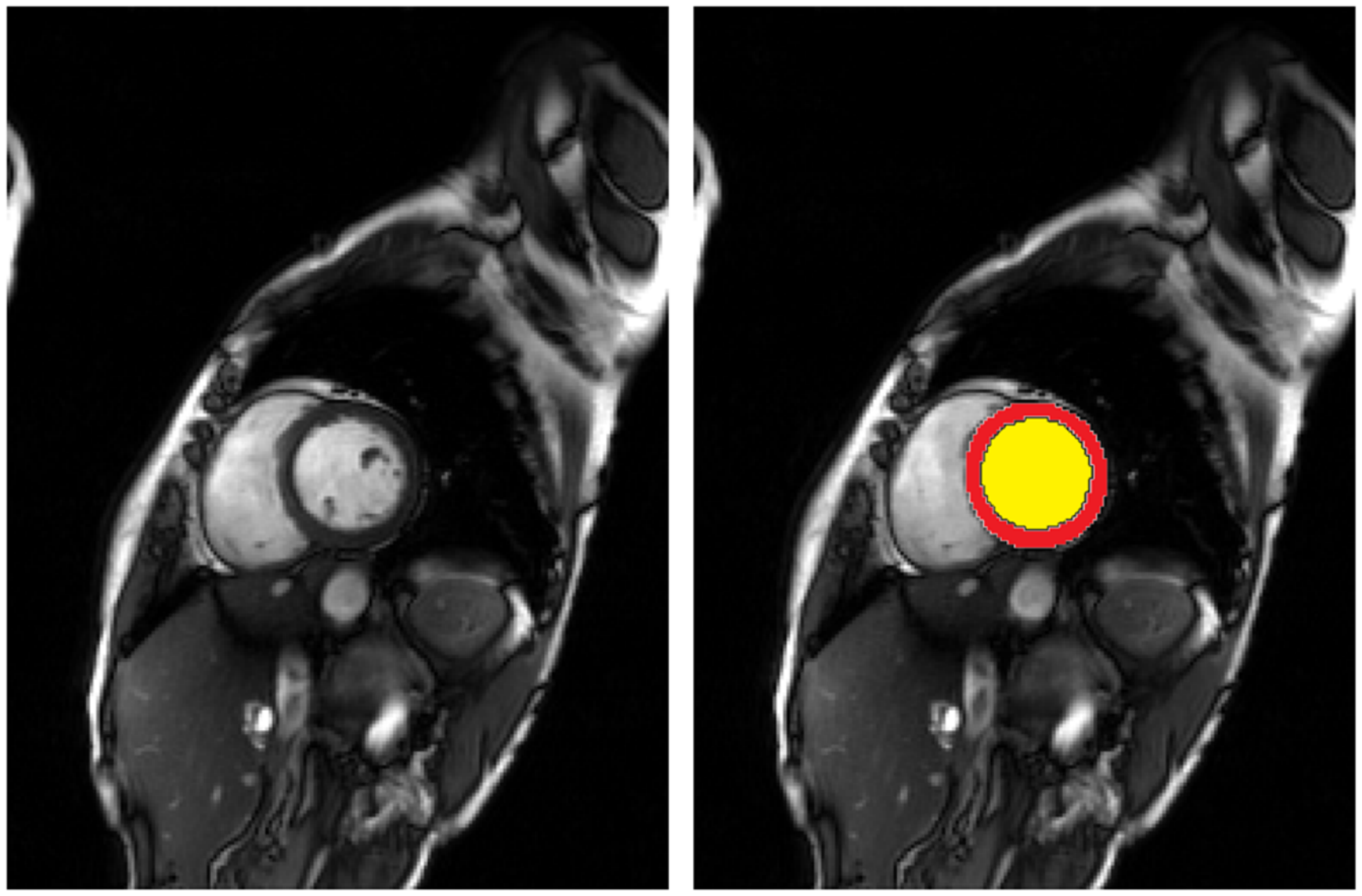

(c) An example of short-axis scan as the ground-truth and segmentation masks.

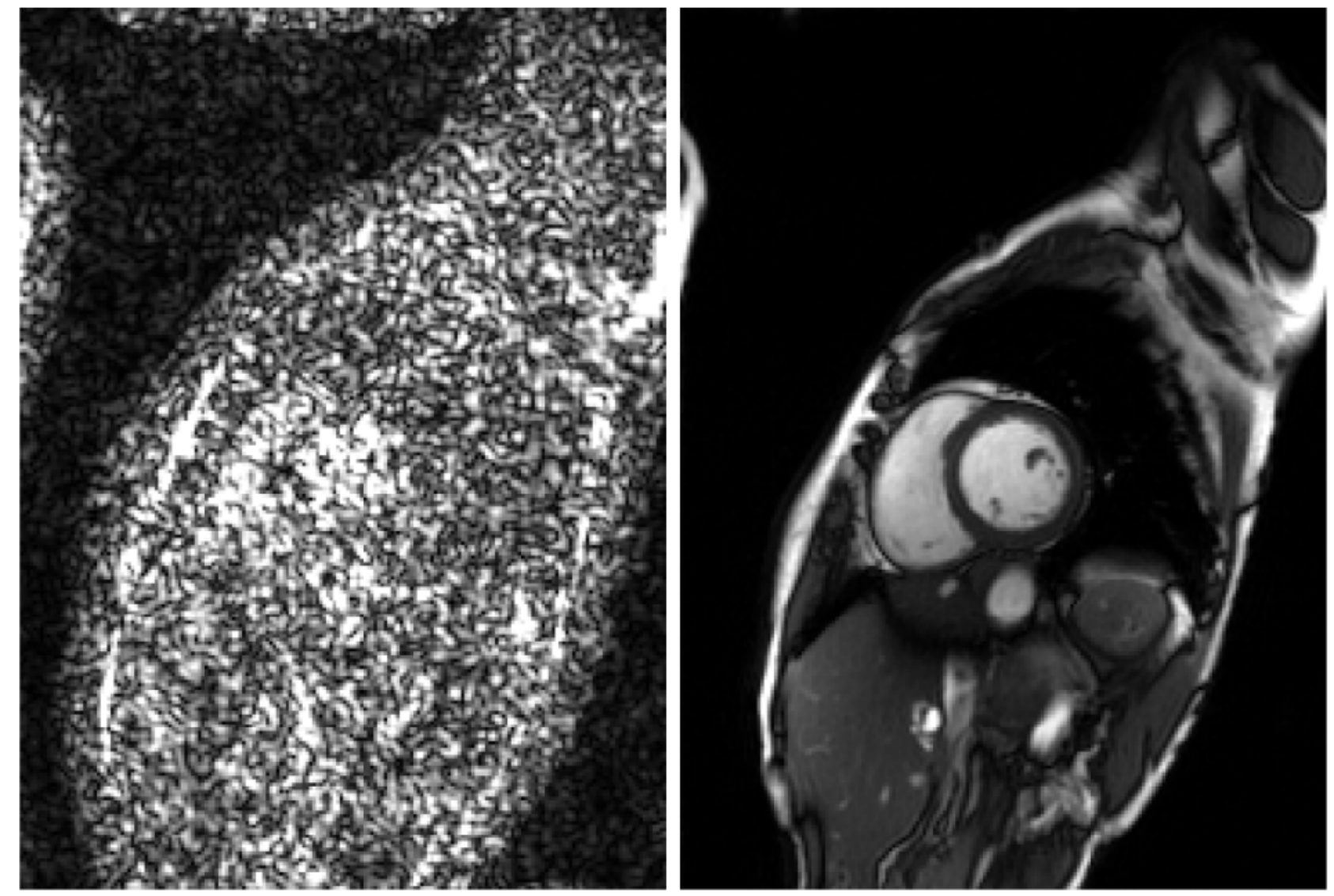

(d) Input median SNR is 0.1 for this case. The blood pool and myocardium contrast was 5.72 and increased to 112.3. The ground-truth CNR is 113.2.

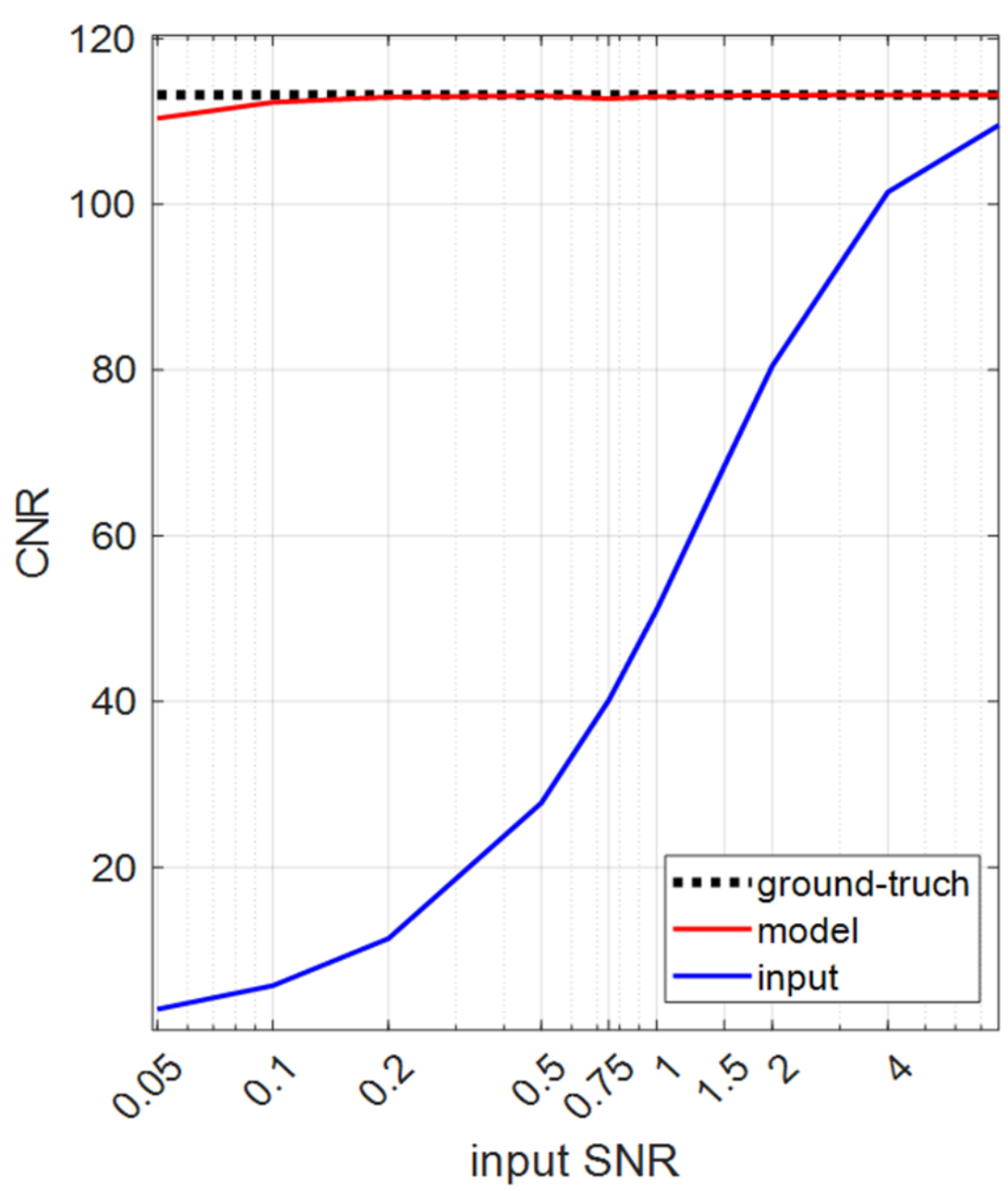

(e) CNRs before and after model are plotted.

## Supplemental Appendices

### Appendix E1. Information for deep learning models

A key design choice is the 5D tensor [B, C, F, H, W]. Because all cells consume and produce the 5D tensor, any number of cells can be inserted into a block. A block also respecting this convention, can be linked together, to compose model backbone.

Different backbone designs are available. Choices include a stack of blocks or simple feed-forward network or adapting the successful CNN backbones and converting them into a transformer model. In this study, we picked the high-resolution network as the backbone. This design maintains a long process branch with the original matrix size, which is suitable for denoising tasks, because the pixel-level prediction is required. In contrast, if the aggregated prediction, such as segmentation or classification, was the target, maintaining original matrix size may not be necessary. In those cases, U-net type backbone with early downsampling can reduce computing cost.

Downsampling was implemented with patch merging (21) followed by a convolution to alter the number of channels if needed. The upsampling was implemented with a linear interpolation followed by a convolution.

Recent study showed the small patch size led to improved performance (51). The minimal patch is a single pixel. It can lead to an attention sequence length of a few thousand. In this study, we chose patch size 2×2 and window size 8×8. For Swin3D and ViT3D, a window includes the third dimension. The 3D window size there was 8×8×16 for H, W, and F.

## Supplemental Figures

### Figure E1. Comparison of EF measurements for small and large models.

The improved image quality from scaling up the model sizes was propagated to the more accurate EF measurements. The Bland-Altman plots for IT-27m were given there. Both the mean error and the 90% CR were much higher than IT-218m model. At SNR 0.2, the EF error was 4.463% for IT-27m, and 0.7975% for IT-218m. The bar plot showed the confidence range where clear gaps were seen at low SNRs.

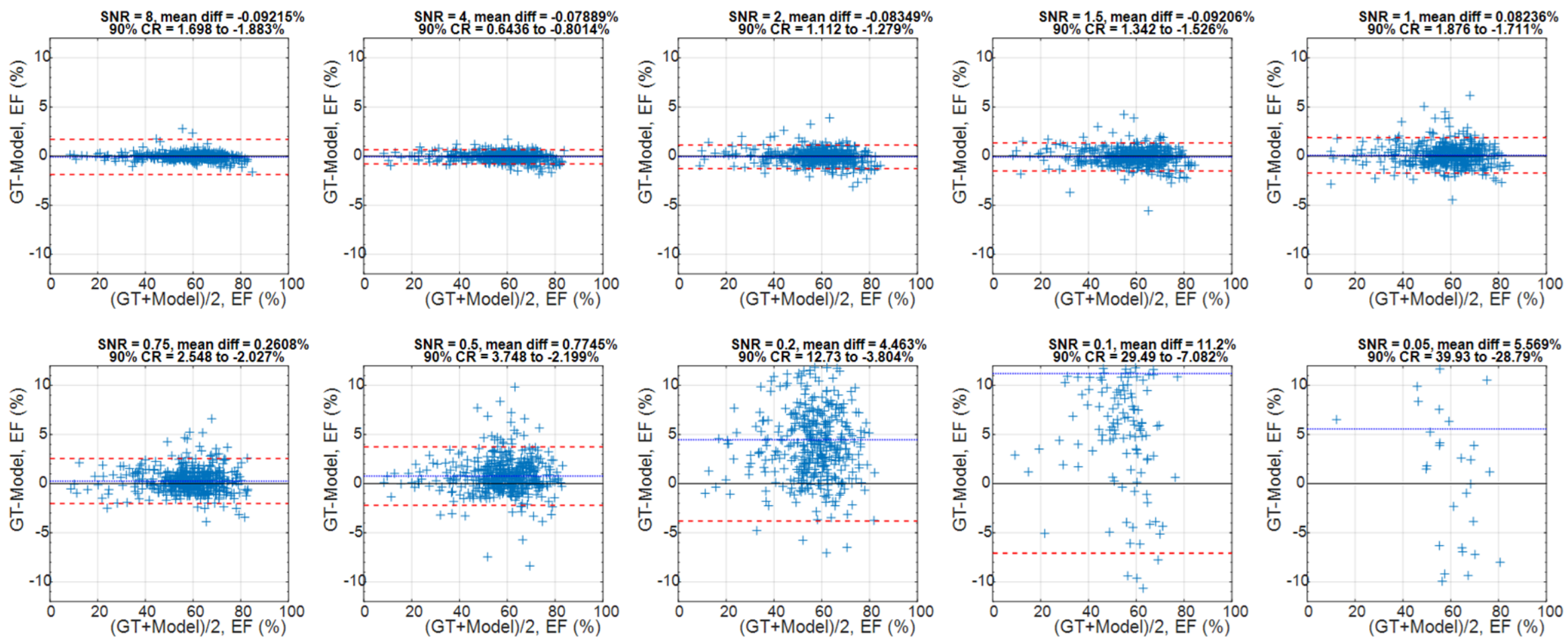

(a) Bland-Altman plots of input SNR vs. EF for **IT-27m**.

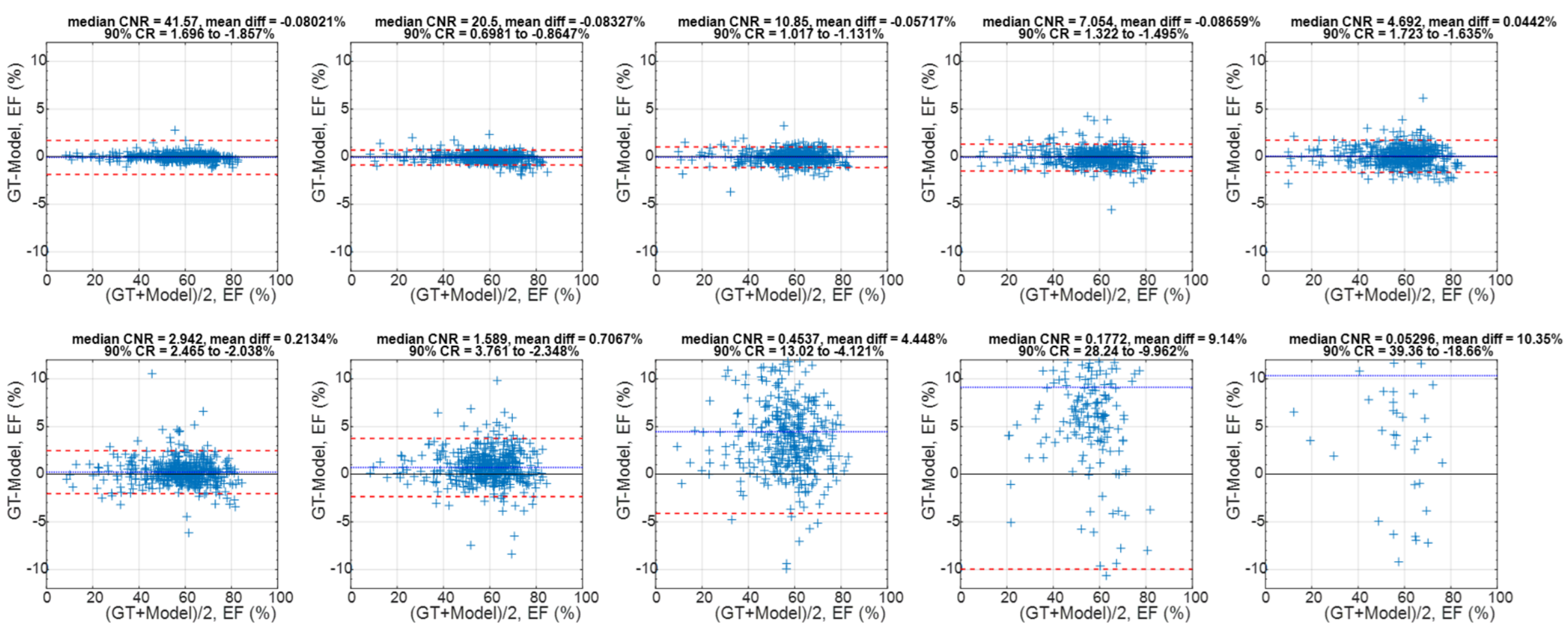

(b) Bland-Altman plots of input CNR vs. EF for **IT-27m**.

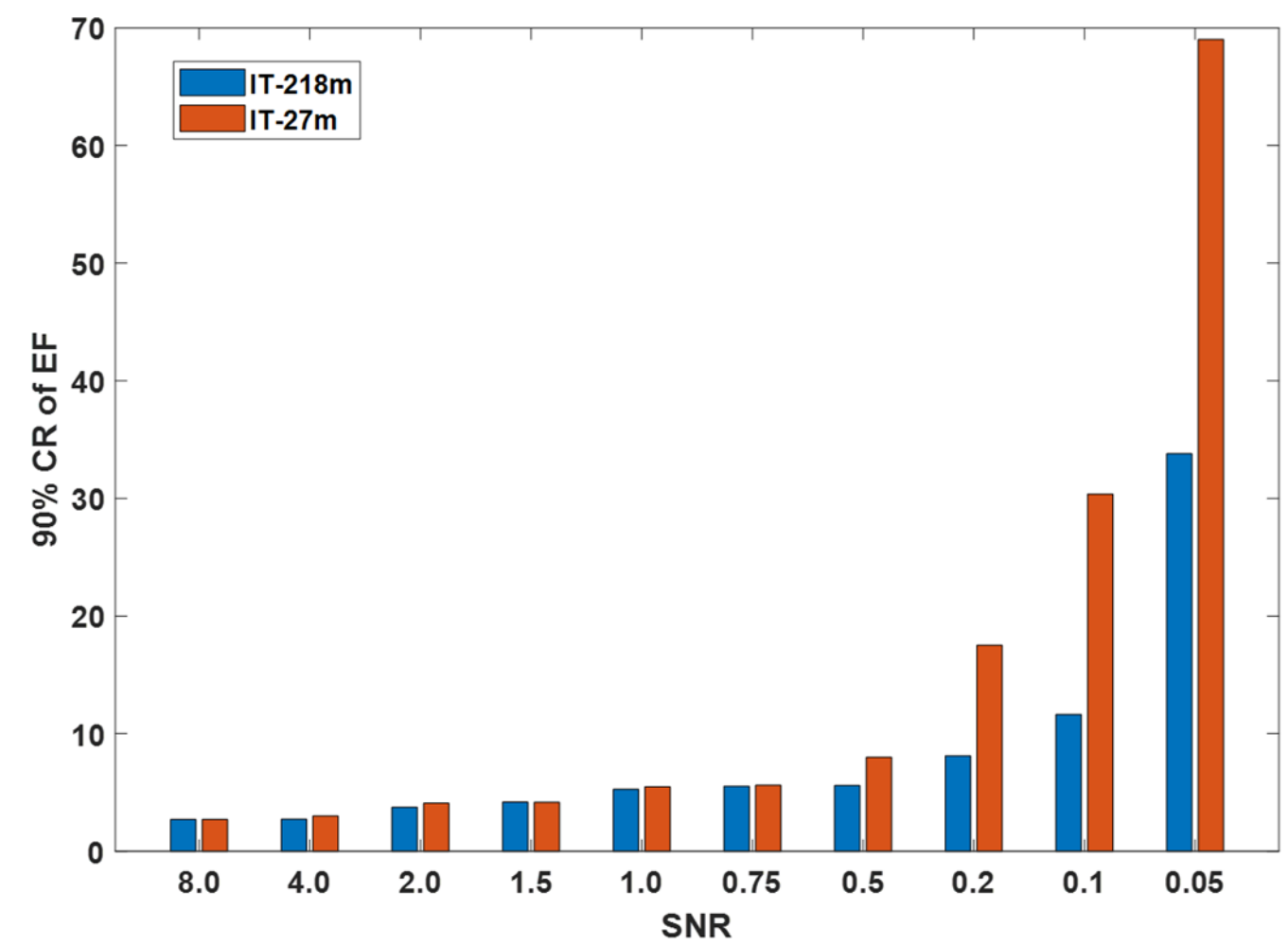

(c) Comparison of 90% CR of IT-218m and IT-27 models for EF.

**Figure E2. Model performance on input data with spatially variant noise amplification from R=2-5 acceleration.**

The model was robust to remove the spatial variant noise due to g-factor amplification. (a) A CH2 cine was corrupted for (b) g-factor maps from R=2 to 5 to create input data with median SNR from ~0.02 to 7.8. (c) Input data to the model at low SNR showed severely degraded image quality and loss of contrast. (d) The corresponding model outputs revealed the limitations of signal recovery. The output quality degrades visibly when input SNR was lower than ~0.2.

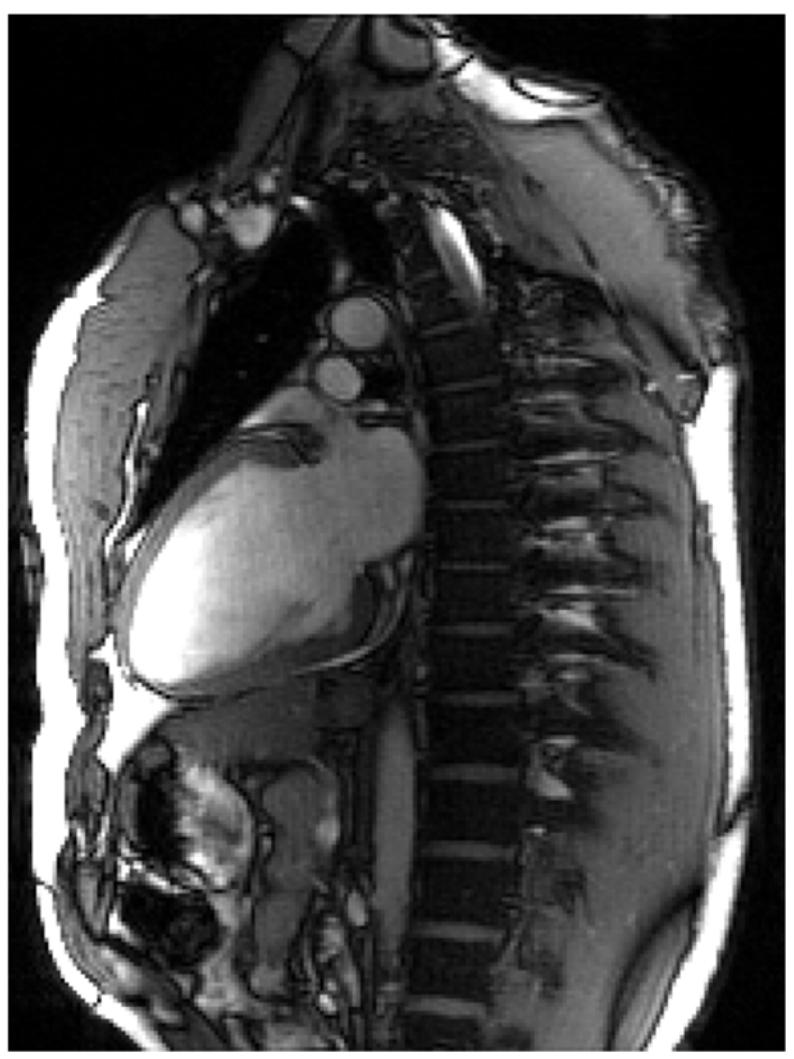

(a) Ground-truth CH2 cine.

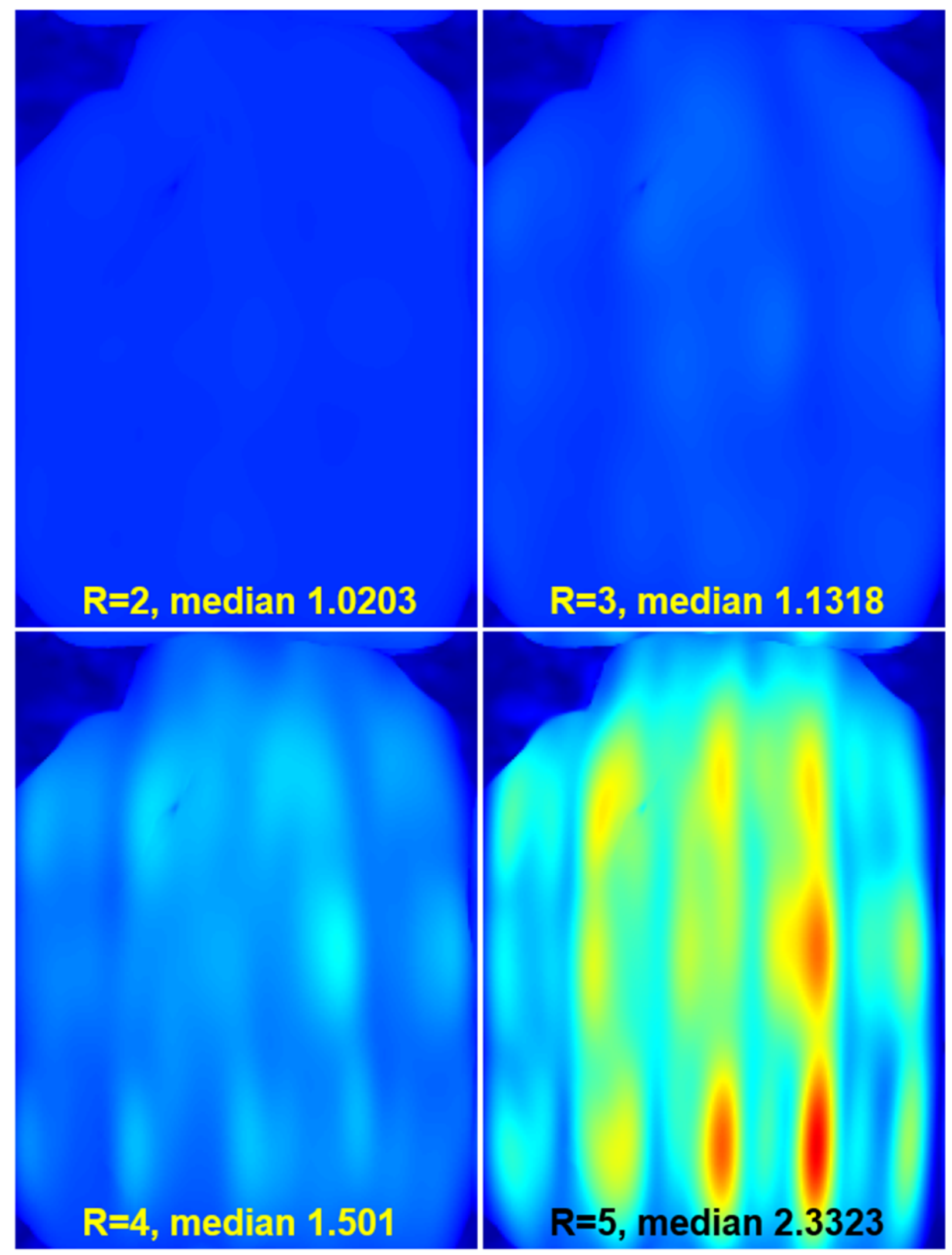

(b) G-factor maps for R=2 to 5, used for generating noise with spatial amplification.

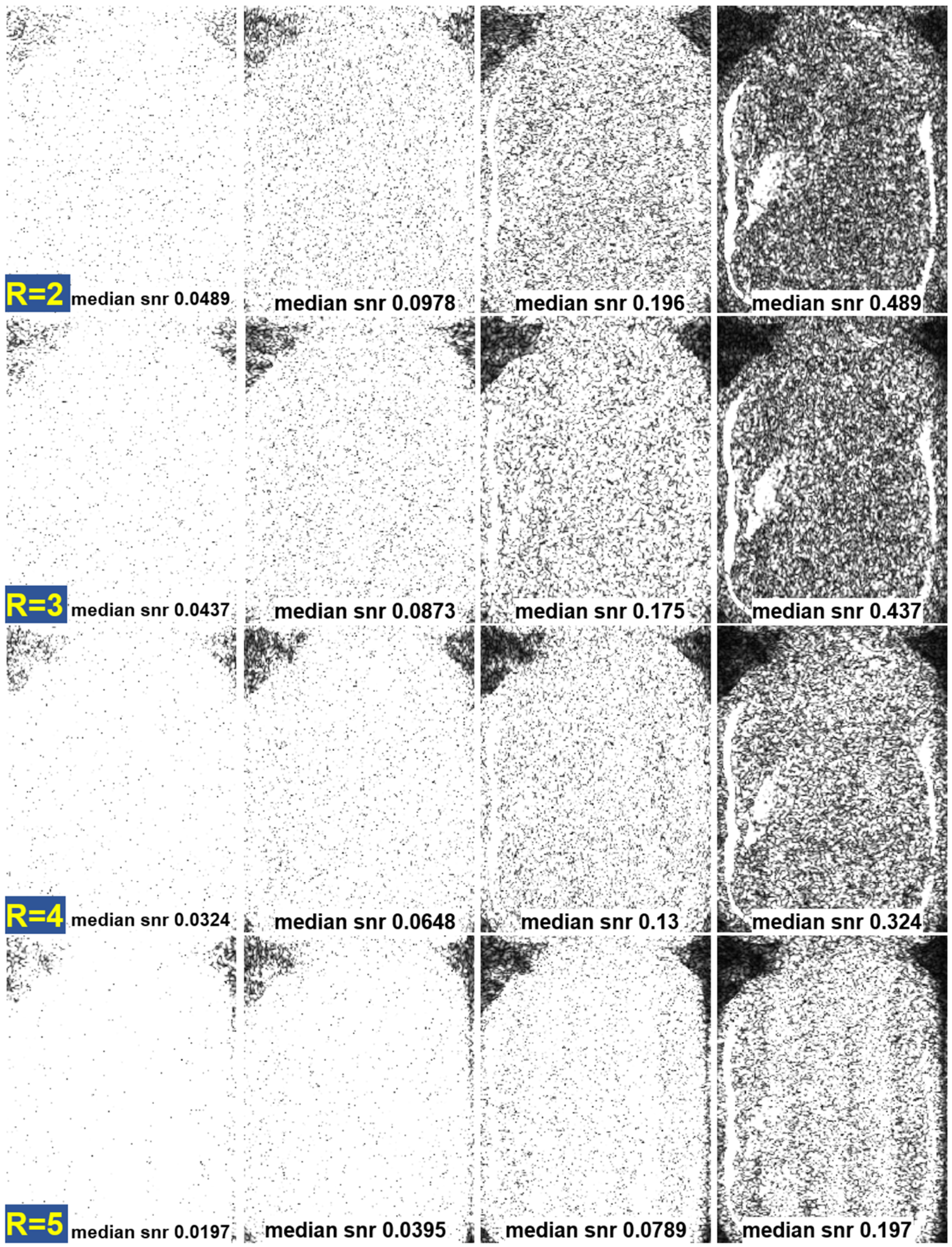

(c) Low SNR inputs for model inference.

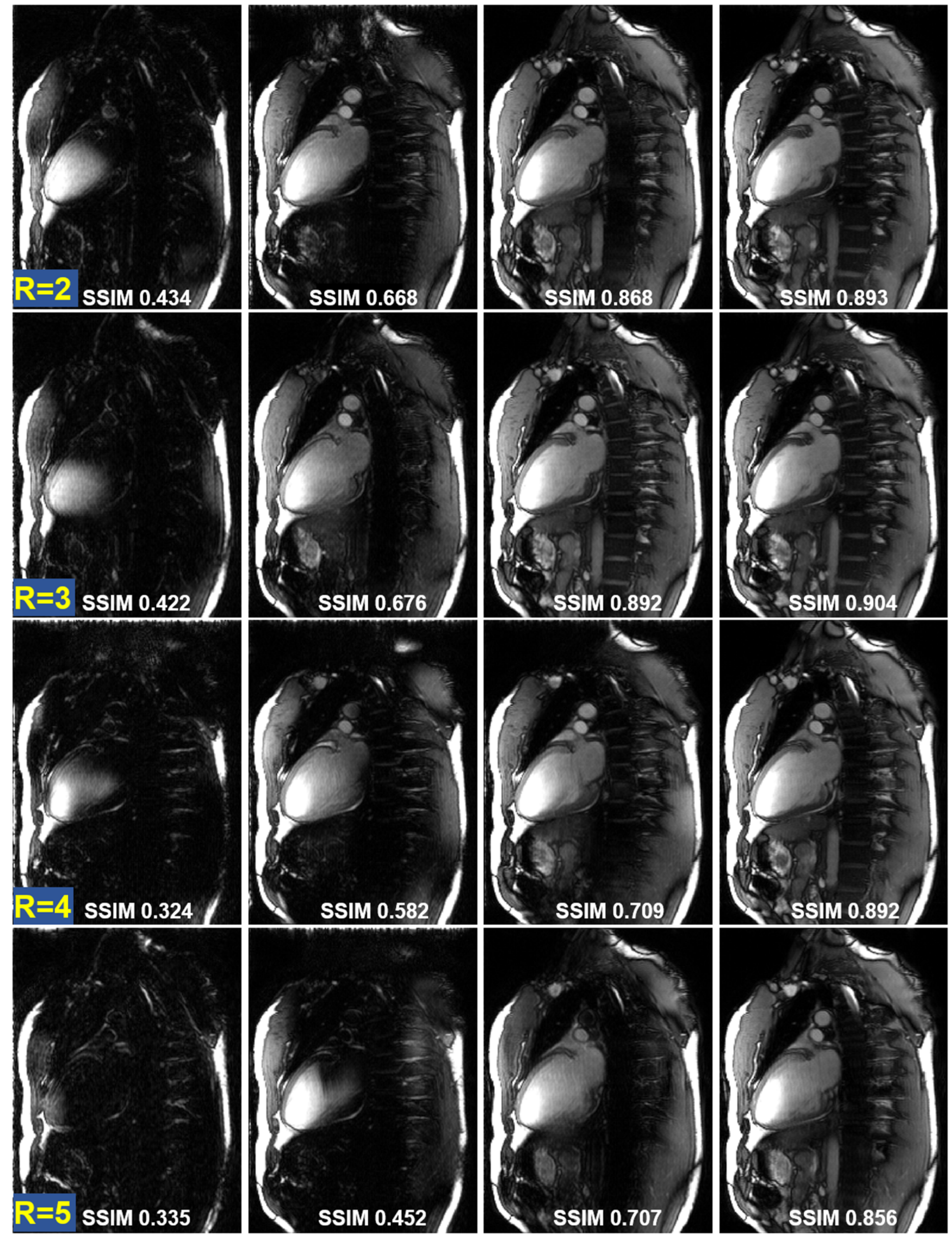

(d) Model outputs. SSIM was computed against the GT.

## Supplemental Data

**Movie 1**: The movies correspond to the example in Figure 2.

**Movie 2**: Examples of ground-truths and model outputs, reviewed by cardiologists. All 11 movies were presented to the reviewers all together. The first one was the ground-truth and others were from the model.

**Movie 3**: The movies correspond to Figure 3. The right panel is the zoomed-in view around the beating heart.

**Movie 4**: The movies correspond to Figure 4. The lower panel is the zoomed version.

**Movie 5**: A four-chamber example of IT-218m model.

**Movie 6**: A short-axis example of IT-218m model.

**Movie 7**: A two-chamber example of IT-218m model.

**Movie 8**: More examples with input SNR level 0.2. The ground-truth is given on the right for reference.

**Movie 9:** The movies correspond to Figure 5. An example of short-axis stack processed by the model for the EF measurements. The pre-trained cine analysis model was applied to both GT and model output SAX stacks. Two EF estimates were computed, independently. If a model faithfully recovered image quality, its EF should agree with the ground-truth EF.

**Movie 10:** The movies correspond to Figure 6 for the blood and myocardial signal level measurement.

**Movie 11:** The movies correspond to Figure E2 (supplement), panel c. A set of examples were created for acceleration R=2 to 5 for SNR from 0.0197 to 7.82. The spatial variant noise was visible at R=4 or 5.

**Movie 12:** The movies correspond to Figure E2 (supplement), panel d. The model processed low SNR data and restored image quality over a wide range of input SNR and spatially variant noise amplification. For input SNR ~0.2 (two examples in red box), the SSIM of model were ~0.85. Further lower SNR led to decay of output quality, showing the current limitations of model.